\documentclass[11pt]{article}
\usepackage[tbtags]{amsmath}
\usepackage{amssymb,multirow,jheppub,graphicx}
\usepackage{centernot}
\usepackage{color}
\allowdisplaybreaks[1]
\usepackage{tikz}
 \tikzset{node distance=2cm, auto}

\usepackage[indent]{parskip}

\DeclareMathOperator{\tr}{tr}

\def\tfrac#1#2{{\textstyle{\frac{#1}{#2}}}}

\def\bar{\overline}
\def\del{{\partial}}

\def\bfn{n}
\def\nt{{\tilde n}}

\newcommand{\im}{\mathrm{i}}

\def\Z{{\mathbb Z}}
\def\R{{\mathbb R}}
\def\coeff#1#2{{\textstyle {\frac {#1}{#2}}}}

\def\half{\coeff 12}

\usepackage{verbatim}   
\usepackage[all]{xy}
\usepackage{bm}  
\def\Dslash{{\rlap{\raise 1pt \hbox{$\>/$}}D}}
\def\Pslash{{\rlap{\raise  1pt \hbox{$\>/$}}\,\partial}}

\newcommand{\lb}{\bm[}
\newcommand{\rb}{\bm]}

\def\a{\mathsf{a}}
\def\b{\mathsf{b}}

\newcommand{\wbar}{{\overline{w}}}

\newcommand{\alphadot}{{\Dot{\alpha}}}
\newcommand{\betadot}{{\Dot{\beta}}}

\sbox0{$\relax$<
\fontdimen16\textfont2=3pt
\fontdimen17\textfont2=3pt   
\fontdimen14\textfont2=4.8pt
\fontdimen13\textfont2=4.8pt>} 

\title{ Self-dual monopole loops, instantons and confinement
  }
\author[1]{Mendel Nguyen}
\emailAdd{mendelnguyen@gmail.com}
\affiliation[1]{Department of Mathematical Sciences, Durham University, Durham DH1 3LP, UK}
\author[2]{Mithat \"Unsal}
\emailAdd{unsal.mithat@gmail.com}
\affiliation[2]{Department of Physics, North Carolina State University, Raleigh, NC 27607, USA}

\abstract{ 
It is well-known that the  standard instanton analysis in 4d Yang-Mills  is plagued with   the instanton size moduli problem, which renders the instanton contribution to vacuum energy density  (or one-instanton partition function)  infrared divergent.    The formalism also ignores the implications of  long range (magnetic dipole type) $1/r^4$ interaction between the small instantons, since it is weaker than Coulomb interaction.   
   We show that  in $U(1)$ lattice gauge theory, where finite action configurations are  monopole  loops,  small loops at large separations also interact with the same type of  $1/r^4$  interaction.  If one ignores the  classical interactions between monopoles, following the same idea as in Yang-Mills theory,  the one-monopole partition function is also infrared divergent  at strong coupling.  However,  $1/r^4$  interactions among small loops should be viewed as a consequence of multipole expansion, and emanate from $1/r^2$   interaction between current segments. 
   Taking interactions into account, one can prove that the strongly coupled $U(1)$ lattice gauge theory is dual to a lattice abelian Higgs model, and more importantly, free of infrared divergences.  
   The model exhibits mass gap and confinement by monopole condensation.  We suggest that the structure of moduli space of instantons, ADHM data,  and the long ranged classical   interactions in pure Yang-Mills theory should be examined with this refined perspective. We conjecture that, in contradistinction to the current views on the subject, internal structure of instantons in Yang-Mills theory is responsible for confinement in $4d$ , similar to sigma model in $d=2$ dimensions.

    }

\begin{document}

\maketitle 

\section{Introduction}

The best understood examples of confinement on infinite $d=4$ dimensional spacetimes are $U(1)$ lattice gauge theory on $\mathbb Z^4$  
\cite{Polyakov:1975rs, Banks:1977cc, Peskin:1977kp, Stone:1978mx, Ukawa:1979yv, Guth:1979gz, Cardy:1981fd} and  ${\cal N}=2$ supersymmetric gauge theory softly broken down to ${\cal N}=1$ (Seiberg-Witten theory) on $\mathbb R^4$ \cite{Seiberg:1994rs, Argyres:1994xh, Klemm:1994qs, Douglas:1995nw, Nekrasov:2002qd}. 
In the latter, in continuum, $SU(2)$ gauge structure is reduced down to $U(1)$ in the long distances. In both cases, abelian duality plays a prominent role and condensation of monopole particles  (or equivalently,  proliferation of the monopole loops on the lattice case) are responsible for confinement.  

Non-perturbative aspects of QCD and  pure Yang-Mills theory are much less understood on $\mathbb R^4$ where dynamics remains non-abelian at all length scales. However, with the use of adiabatic continuity and compactification  
 on  $\mathbb R^3 \times S^1$  
\cite{Unsal:2007jx, Unsal:2008ch} (in constrast to the  thermal compactification)\footnote{Thermal compactications generically possess phase transitions, but is possible to construct nonthermal compactifications without them. This idea is called adiabatic continuity. See \cite{Dunne:2016nmc, Poppitz:2021cxe} for a review. } 
and $\mathbb R^2 \times T^2$ with 't Hooft flux \cite{Tanizaki:2022ngt},   
 confinement, mass gap and fractional theta angle dependence are generated by 
 fractional instantons. These are monopole-instantons  and magnetic bions  on $\mathbb R^3 \times S^1$ and  center-vortices on $\mathbb R^2 \times T^2$.


It is clear that  $U(1)$ lattice gauge theory on $\mathbb Z^4$  and  $SU(N)$ Yang-Mills theory in continuum  $\mathbb R^4$ are very different theories. The former is on the lattice, abelian, with a tunable bare lattice coupling, and the latter  is in the continuum, non-abelian at all length scales, with an untunable coupling that exhibits asymptotic freedom. Apparently, they do not seem to do much with each other.

Although many aspects of $U(1)$ lattice gauge theory are well known,  in our opinion, the most important lessons emanating from it that  could potentially be useful to resolve some longstanding problem in Yang-Mills theory are overlooked. In the standard semiclassical treatment of Yang-Mills theory, there is a notorious instanton size modulus problem, the fact that the  contribution of large-instantons to vacuum energy density diverges. This is also known as the ``Infrared embarassment'' \cite{Coleman:1985rnk}.  This problem is solved on  $\mathbb R^3 \times S^1$  \cite{Unsal:2007jx, Unsal:2008ch}.  However, on $\R^4$,  
truth is no one knows what to do with the large instantons.    
Below, we will show that the exact counterpart of this problem is also present in $U(1)$ lattice gauge theory on $\Z^4$.  In particular, the integral over instanton size is replaced with the integral over the monopole loop size, both of which are divergent in strongly coupled regimes. 
Yet, as we will prove rigorously, taking interactions into account, the theory resolves the problem in a satisfactory manner.  The solution gives strong hints on how to work  with instantons in Yang-Mills theory.


\subsection{Aspects of  $U(1)$ lattice gauge theory}

\begin{figure}[t]
\begin{center}
 \includegraphics[angle=0, width=0.7\textwidth]{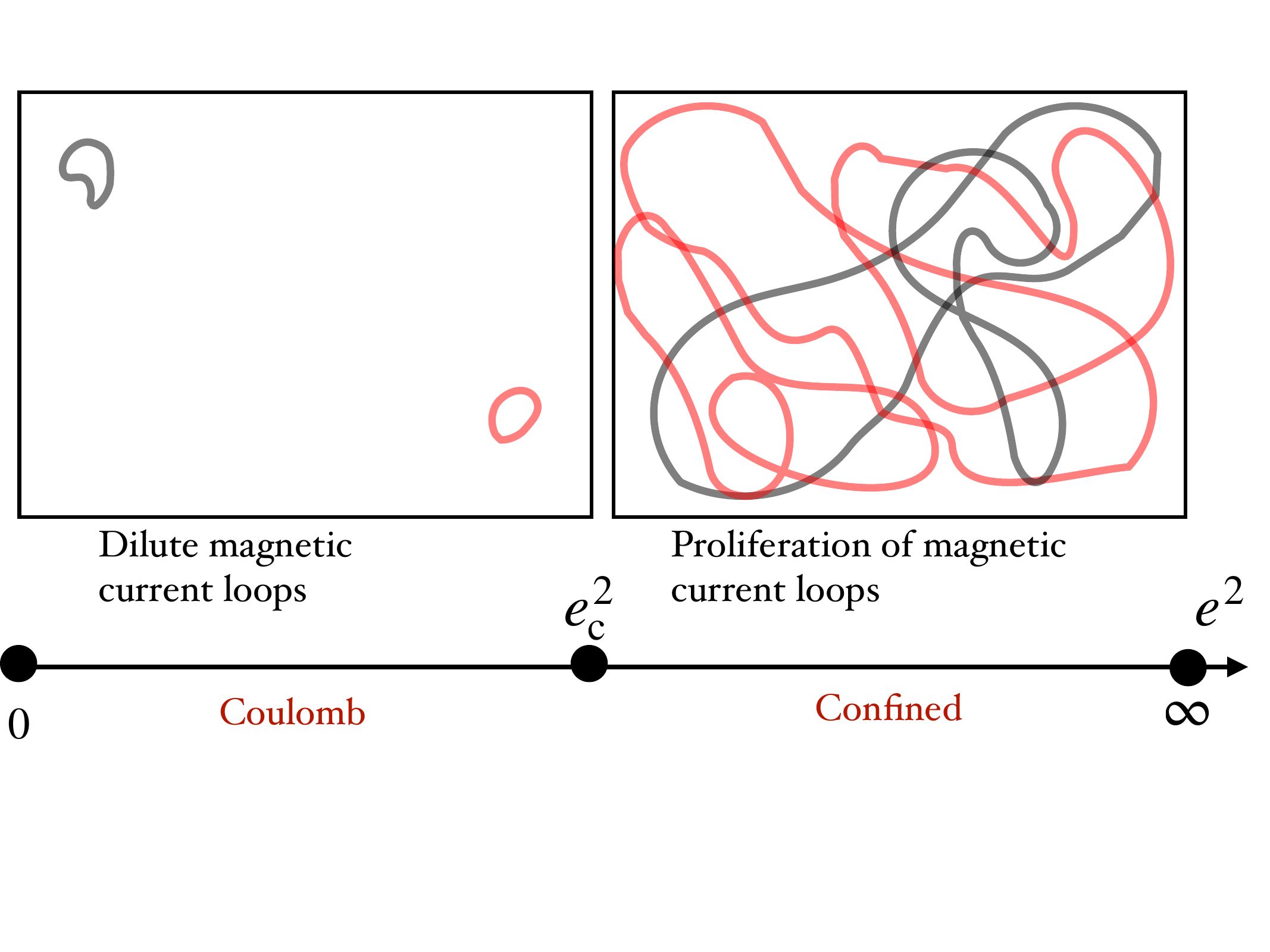}
\vspace{-2cm}
\caption{$U(1)$ lattice gauge  in 4d theory has two phases.  A weak bare coupling Coulomb phase in which monopole current loops are rare and small, 
 a  strong bare coupling confined phase in which monopole current loops  proliferate and large.      }
\label{phasesu1}
\end{center}
\end{figure}

 Pure   $U(1)$ lattice gauge theory in $d=4$ has two phases \cite{Polyakov:1975rs,Banks:1977cc}, see Fig.\ref{phasesu1}a).  
A strongly coupled  (latticy) confining phase for $e^2  > e_{\rm c}^2$ \cite{Wilson:1974sk}, and a weak coupling Coulomb phase for $e^2  < e_{\rm c}^2$.  
Microscopically, the theory has monopoles, and the statistical field theory can be viewed as a theory of 
magnetic current loops. 
There is an heuristic explanation based on Peierls instability type argument which suggests that a phase transition ought to occur as a function of bare lattice coupling as one moves from small bare coupling to large.  
A monopole loop of length $L$  
has an action proportional to the product of the monopole mass  $M_{\rm mon} =M$ and the length of the loop $L $, $S_{\rm mon}= M L =  {2 \pi^2} G_0L/e^2  $.
As such,  the  Boltzmann weight of a single such configuration  is  $e^{- M L} $ which implies that the long loops are exponentially suppressed. However,  the  entropic factor  for the  non-backtracking  loops of   length $L$ (that start and end at a fixed point $x \in \mathbb Z^4$)
is given by $e^{ C L}, \; C \sim 1 $   hence the weight factor  for the loops of length $L$ is given by 
\begin{align}
Z^{(1)}_L =  e^{-  M L}   e^{ C L},   \qquad   M =  \frac{2 \pi^2}{ e^2} G_0
\end{align}
At strong bare coupling $e^2 > e_{\rm c}^2$ regime , the entropy dominates over the energy and 
$Z^{(1)}_L$ grows with $L$, and 
loops proliferate.    At weak bare coupling, the energy dominates over entropy, and large monopole loops are suppressed.  This is a well-known story.

If we ignore interactions between current loops,  we can express the  1-monopole contribution to  partition function by summing over all configurations of  the monopole loop (in analogy with 1-instanton contribution to partition function in Yang-Mills theory, see \cite{Callan:1977gz, Polyakov:1987ez, Coleman:1985rnk} in the so called dilute instanton gas 
) as:
 \begin{align}
Z^{(1)}_{\rm mon}  &= \sum_{x \in  \mathbb Z^4 } \sum_{L} e^{- M  L}    e^{ C L}  
\sim {\rm Vol} ( \mathbb Z^4) \int \frac{dL}{L} \;  e^{- M  L}    e^{C L}   
\label{1-monPF}
\end{align} 
 This {\it would} imply that the contribution of monopole loops to vacuum energy density is IR-divergent  in the strong coupling  $e^2 > e_{\rm c}^2$ regime, just like similar contribution of 1-instanton \eqref{DIG}. However, this problem is fake, 
 it is an artifact of our blatantly incorrect approximations, neglecting the interactions between the monopole current loops.   
As we will describe after we review instantons in Yang-Mills theory, this divergence is the  counterpart of instanton size problem.  
Below, we outline how lattice gauge theory resolves this problem.

\begin{figure}[t]
\vspace{-1cm}
\begin{center}
 \includegraphics[angle=0, width=0.8\textwidth]{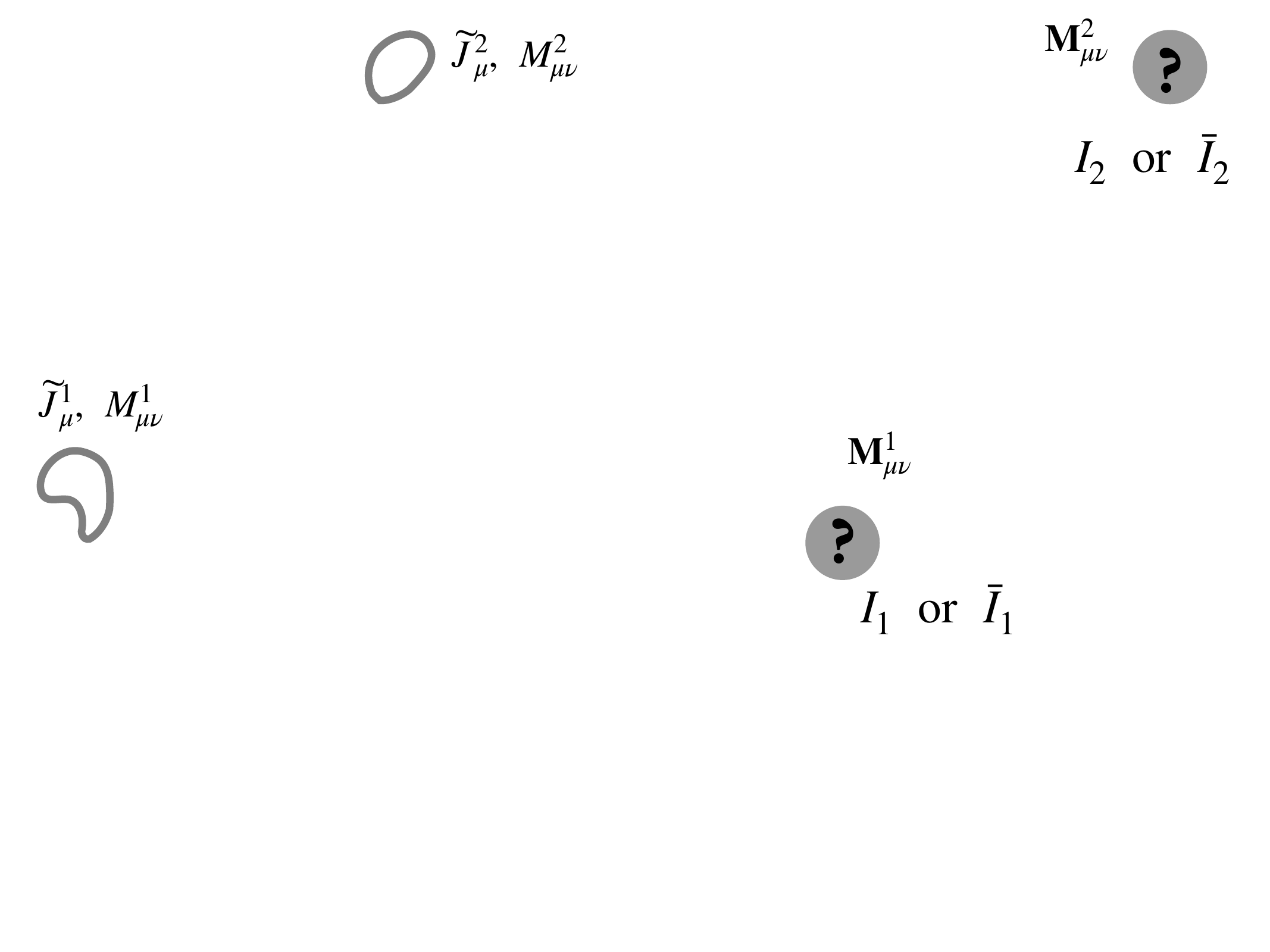}
\vspace{-3cm}
\caption{a) Small magnetic current loops  interact via  magnetic dipole interactions decaying algebraically as $1/|x|^4$ at long distances \eqref{magic0}. b) Remarkably, small instantons in $SU(N)$ gauge theory also interacts via a (non-abelian) magnetic dipole type interactions at large separations \eqref{magic2}. Why is this so, and what is the internal structure of instantons which leads to this interactions? 
     }
\label{currentloops}
\end{center}
\end{figure}

The resolution comes from  the inclusion of interactions between current loops.  However, to 
draw a close analogy with the idea of instanton, we will first talk about the interaction between small loops at large separation (similar to small instanton interactions at large separation \cite{Callan:1977gz}).
 Their mutual interaction can be expressed in terms of their magnetic dipole moments  $M_{\mu\nu}, \mu,\nu=1, \dots, 4$, a  second-rank anti-symmetric tensor.  The interaction between two small loops (shown in Fig.\ref{currentloops}) at a large separation is given by:
 \begin{align} 
   V_{\rm  loop-loop}^{\rm mon}  &=  
   \frac{C}{e^2}  \frac{1}{|x|^4}   \left( 4 ( M^{1}_{\mu \nu}  \hat x_\mu )(   M^{2}_{\alpha \nu}    \hat x_\alpha ) -   M^{1}_{\mu \nu}  M^{2}_{\mu \nu}  \right)  
   \label{magic0}
\end{align}
In the  $e^2 < e^2_{\rm c}$ regime, the effect of the small loops just renormalizes 
gauge coupling for the Maxwell term. As such, they cannot change (in the dilute regime)  the Coulomb interaction between external probes into a confining one.  Based on this lack of 
qualitative impact on the dynamics on the weak coupling regime, and the fact that the interaction between small loops is  ${1}/{|x|^4}$, 
one may be tempted to ignore the interactions altogether. (This is essentially what is done in the dilute instanton gas ``approximation'' because interaction between small instantons is 
${1}/{|x|^4}$.) 
However, this would be a grave mistake, because in the regime 
where loops are non-dilute, the interaction between current bits is of the form  $1/|x|^2$, and can alter the behaviour of the theory drastically.  In particular, ${1}/{|x|^4}$ interaction ought to be thought as arising from multipole expansion of small size loops. The underlying interaction in the case of lattice gauge theory is $1/|x|^2$ and cannot be ignored.

We want the reader to leave this short overview with two main observations: 
\begin{itemize}
    \item Small monopole loops at  a large separation   interact via an algebraic $1/|x|^4$ dipole-dipole interaction. 
    \item At  strong bare lattice  coupling $e^2 > e^2_{\rm c}$,  one-monopole contribution to partition function diverges  rather badly at large-$L$ by large monopole loops.  
\end{itemize}

As the reader may anticipate, we isolated these two important facts for a reason. As we summarize below, the counterpart of these two in Yang-Mills theory and QCD are exactly the reasons why the QFT community  gave up instantons as a possible mechanism  of confinement on $\mathbb R^4$ as most clearly stated by Coleman and 
Polyakov\cite{Polyakov:1987ez, Coleman:1985rnk}.\footnote{Coleman in  ``The uses of instantons'', after an explanation states: ``Whatever confines quarks, it is not instantons.''
\cite{Coleman:1985rnk}. 
Polyakov, in his memoirs,  on the issue of non-abelian confinement states that, ``Here the
instantons disappointed me.''\cite{Polyakov:2004vp}. This is not a criticism to neither author, 
to the contrary,  we sincerely appreciate their frankness.}
Yet, for $U(1)$ lattice gauge theory on $\mathbb Z^4$, these two features are present exactly in the same way,  but as we show below,  they do not constitute a problem. The reason is,  in the strongly coupled  phase in which magnetic  loops proliferate, and the naive dilute loop gas description leads to an IR divergence, 
we can actually rigorously dualize the system to 
a magnetic abelian Higgs model, and prove  that 
there is no IR-divergence, the system is gapped  out, and  linear confinement sets in by  monopole condensation.

\subsection{Aspects of  $SU(N)$ Yang-Mills theory}

The Yang-Mills theory on $\mathbb R^4$  possess  finite action non-perturbative saddles, called instantons.  These are, unlike monopole particles in $U(1)$ gauge theory  which are represented as worldlines in $\mathbb R^4$, are events in space-time.  
As such, they do not resemble much to each other, but on the other hand, both monopole loops and instantons have finite actions.

Instantons in 4d has a rather large classical moduli space. For one-instanton in the $G=SU(N)$ gauge theory,  the dimension of moduli space is $ {\rm dim} ({\cal M})= 4N $ \cite{Dorey:2002ik}.
There are different useful ways to parametrize the moduli space. One is $X \in \mathbb R^4$ is the center,   $\rho \in \mathbb R^{+}$ is the size modulus, and angular moduli that live in a $4N-5$ dimensional compact coset space:  
\begin{align}
  {\cal M} =  \mathbb R^4  \times   {\cal M}_{\rm c} =  \mathbb R^4 \times \mathbb R^+ \times   {\widehat {\cal M}}_{\rm c},  \qquad \widehat {\cal M}_{\rm c} =  \frac{SU(N)}{S( U(N-2) \times U(1))}
  \label{moduli}
\end{align}
where ${\cal M}_{\rm c}$ is called centered moduli space, and $\widehat {\cal M}_{\rm c}$ is compact coset.\footnote{Instantons, in classically non-scale invariant theories, do not have the counterpart of size moduli, hence deserved to be called ``events,'' localized in space-time. 
However, in classically scale invariant theories, the size of an instanton is a modulus that one integrates over. Then, it becomes questionable to what extent this configuration should be viewed as localized event in spacetime. It may very well be filling or exploring the whole space. 
Therefore, especially in classically scale invariant theory, one should not be dismissive about the classical interactions of instantons, as it may reveal inner structure thereof.  
For application of this idea to $\mathbb {CP}^{N-1}$, see  \cite{Nguyen:2023rww}.}
In certain occasions, a more useful parametrization is the one that appears in ADHM formalism \cite{Atiyah:1978ri,Dorey:2002ik}.  We employ both.

 It is a known, but not a widely utilized fact that the small instantons in pure Yang-Mills theory at a large-separation, see Fig.\ref{currentloops}, interact via a {\it classical} potential: 
  \begin{align} 
   V_{\rm int}^{\rm inst}  & =\frac{4 \pi^2}{e^2}  \frac{1}{|x|^4}   \left( 4 ( M^{1a}_{\mu \nu}  \hat x_\mu )(   M^{2a}_{\alpha \nu}    \hat x_\alpha ) -   M^{1a}_{\mu \nu}  M^{2a}_{\mu \nu}  \right)  
   \label{magic2}
\end{align}
For $SU(2)$, this interaction was derived in \cite{Callan:1977qs}. For some phenomenological applications, see \cite{Schafer:1996wv}. 
 The reader (as the authors) should feel quite surprised by this result.  The main 
puzzling question is:
\begin{itemize}
    \item Why the interaction between instantons \eqref{magic2} is  exactly of the form of two 
     magnetic dipoles \eqref{magic0} (with the replacement of abelian magnetic dipole with non-abelian self-dual dipole)?  Can we use this fact to learn about the underlying structure of an instanton? 
\end{itemize}


It is not completely sensible to view a non-abelian instanton as a simple magnetic dipole of some size $\rho$.\footnote{If the gauge group is broken down to its   abelian subgroup, eg. $SU(2) \rightarrow U(1)$ \cite{GarciaGarcia:2025uub},  or similarly, if one uses the so called maximal abelian gauge as in \cite{Brower:1996js}, there may be precise relations between abelianized instantons and magnetic current loops.} 
It is a much more refined object. In fact, in the full non-abelian theory,  it is a collection of $N^2 -1$ self-dual (or anti-selfdual) magnetic dipoles ($N \times N $ traceless matrix), and hence non-abelian in its nature.  
\begin{align}
&  M^{a}_{\mu \nu} T^a \equiv {\bm M}_{\mu \nu},\; a=1, \ldots, N^2-1, \qquad  \cr 
 & {\bm M}_{\mu \nu}  = \pm \tilde {\bm M}_{\mu \nu}  
\end{align}
However, it is important to note that we do not have $3 (N^2 -1)$ parameters in non-abelian $SU(N)$ gauge theory   which dictate these matrices. Rather, the non-abelian magnetic dipole moment matrix, however, is determined by only $4N-4$ 
of the moduli  parameters of an instanton  parametrizing centered moduli space 
\begin{align}
w_{u \dot \alpha } \in {\cal M}_{\rm c}, \qquad u=1, \ldots, N, \; \dot \alpha=1,2 
\end{align}
 and entering ADHM data \cite{Dorey:2002ik}. Remarkably, the ADHM parameters  $w_{u \dot \alpha }$   and non-abelian magnetic dipole moment matrix is related in a  simple way:
\begin{align}
  {\bm M}_{\mu \nu}  = w \bar \sigma_{\mu \nu} w^\dag  
\end{align}
where $w$'s satisfy the ADHM constraint: $w^\dag w = \rho^2 1_2$, and $\rho$ is identified with the instanton size.  

In semiclassical treatments of QCD and Yang-Mills theory, there is a rather important  ``infrared embarrassment problem'' \cite{Callan:1977gz, Coleman:1985rnk, Polyakov:2004vp}, and the culprit is 
usually viewed as the integration over scale size parameter $\rho \in \R^+$ of instanton.  
One instanton contribution to partition function (hence vacuum energy density)  is proportional to:\footnote{We drop the volume of the compact coset, ${\rm Vol}(\hat {\cal M}_{\rm c})$, which is finite. }
\begin{align}
 Z_{\rm inst}^{(1)} &=  \int d^4X \int  \frac{d \rho}{\rho^5} e^{- \frac{8 \pi^2}{e^2 (\rho)} + \im \theta }   
\; =  \;  {\rm Vol}(\mathbb R^4) \int  \frac{d \rho}{\rho^5}  (\Lambda  \rho)^{\frac{11N}{3}}  e^{ \im \theta } =  {\rm divergent} 
 \label{DIG}
\end{align}
Clearly, in Yang-Mills theory, since the theory becomes strongly coupled in IR, 
we can convert $e^{- {8 \pi^2}/{e^2 (\rho)}}$ to $(\Lambda  \rho)^{\frac{11N}{3}}$  by using dimensional transmutation, and the integral is divergent because of its IR  end, $\rho \rightarrow \infty$. 
However, if for some reason, for example, like Higgsing with fundamental scalars, (which freezes coupling constant at some high scale) we turn off the running of the coupling, the integral  becomes  convergent, and dilute gas becomes meaningful. 


As we described in $U(1)$ lattice gauge theory, the same problem is also present in the ``non-interacting  monopole loop gas,'' and there,  the same problem is actually fake,  an artifact of not taking interactions between magnetic currect loops into account.  
The issue of divergence is actually settled if we careful take into account the interaction between current segments. In the strongly coupled  regime, the theory can be dualized to an abelian Higgs model at weak coupling  (similar to Seiberg-Witten theory at monopole or dyon points \cite{Seiberg:1994rs}) and monopole condensation sets the vacuum structure.  A description of this sort is what we are after for the proliferation of the instantons in Yang-Mills theory.



 \subsection{The use of adiabatic continuity for the analysis on $\mathbb R^4$}
The idea of continuously  connecting a strongly coupled gauge theory on  $\mathbb R^4$ to a weakly coupled regime without any intervening phase transition is called adiabatic continuity 
\cite{Dunne:2016nmc, Poppitz:2021cxe}.
 In recent years, such  constructions on  compactified geometries  have been a very fruitful  playground in understanding non-perturbative phenomena in Yang-Mills and QCD-like theories. 
 Many properties of the 
  theories satisfying adiabatic continuity  have been  investigated vigorously on $\mathbb R^3 \times S^1 $ 
 \cite{Unsal:2007vu, Unsal:2007jx, Unsal:2008ch, Shifman:2008ja,  Argyres:2012ka,  Poppitz:2021cxe, Poppitz:2008hr, Poppitz:2011wy, Poppitz:2012nz, Poppitz:2012sw, Aitken:2017ayq, Cherman:2016hcd, Cherman:2016jtu, Cherman:2017dwt, Aitken:2018mbb, Anber:2013doa, Anber:2014lba}. 
 The semi-classical  studies on $\mathbb R^2 \times T^2 $ with 't Hooft flux satisfying adiabatic continuity is more recent \cite{Tanizaki:2022ngt, Tanizaki:2022plm, Hayashi:2023wwi, Hayashi:2024new, Hayashi:2024qkm, Hayashi:2024yjc, Hayashi:2025doq}. 
There are two very important lessons concerning 4d instantons emanating from  semi-classically calculable compactified set-ups on $\mathbb R^3 \times S^1 $  and $\mathbb R^2 \times T^2 $ adiabatically connected to $\R^4$.


\begin{itemize}
    \item If the interactions between instantons is {\it long-ranged},  it is fundamentally incorrect to be dismissive about their mutual interactions. For example, in Polyakov model on $\mathbb R^3$ \cite{Polyakov:1975rs}, and calculable gauge theories on $\mathbb R^3 \times S^1$ \cite{Unsal:2007jx, Unsal:2008ch} (such as QCD(adj) or deformed Yang-Mills) where confinement is caused by monopole instantons or bions, if one ignores the Coulomb interactions $(1/|x|)$, one loses  both the mass gap and confinement.

    \item In different semi-classically calculable occasions, an instanton fractionates to 
    $N$ constituents. On $\mathbb R^3 \times S^1$, it fractionates to $N$ monopole-instantons (with long-range interactions), and on  $\mathbb R^2 \times T^2$, to $N$ center-vortices (with short-range interactions due to  't Hooft flux).  Whatever theory of instantons we develop on $\R^4$ {\it must} incorporate  the description  on $\mathbb R^3 \times S^1$ and on $\mathbb R^2 \times T^2$ by adiabatic continuity .
    
\end{itemize}

The precise mathematical relation between monopole mechanism on  $\mathbb R^3 \times S^1$ and 
center-vortices on  $\mathbb R^2 \times T^2$  in these semi-classically calculable  regimes is already proven in \cite{Hayashi:2024new, Guvendik:2024umd}.  In the respective set-ups, these configurations are instantons localized in spacetime. 

Furthermore, there has been important progress clarifying the role of monopoles (loops) and center-vortices (surfaces) directly in 4d $\mathbb Z_N$ lattice gauge theory \cite{Nguyen:2024ikq}.  We wish to speculate that instantons, monopoles and center-vortices in 4d gauge theories are not competing scenarios. Rather, monopole and center-vortex like structures must be built into the 4d instanton.  We imagine that this is how they reappear 
once the circumstances demand it to be so: 
monopoles appear as fractional   instantons when  the theory is abelianized down to $U(1)^{N-1}$  on   $\mathbb R^3 \times S^1$ and  center-vortices  appear as fractional instantons  when the theory is reduced further down to  $\mathbb Z_N$ gauge theory due to 't Hooft flux on   $\mathbb R^2 \times T^2$.  We believe that the  monopole type and center-vortex type fractional instantons must be in the internal structure of 4d instantons, already on infinite space in some way. Once compactified, this connection on $\mathbb R^3 \times S^1$ is already known \cite{Kraan:1998sn, Kraan:1998pm, Lee:1997vp, Lee:1998bb, Lee:1998vu}.  

\section{U(1) lattice gauge theory}
\newcommand{\fdif}{\Delta}
\newcommand{\bdif}{\nabla}

As we have mentioned, the case of $U(1)$ lattice gauge theory represents perhaps the best understood model of confinement in four dimensions.
Here, we will review this old story \cite{Polyakov:1975rs,Polyakov:1987ez,Banks:1977cc,Peskin:1977kp,Stone:1978mx}; 
needless to say, we will not uncover anything new about the phase structure of this model.  
What is important and new in our analysis is that, if we analyze the monopole loop gas 
exactly as in the so called dilute instanton gas in Yang--Mills theory, we will face the very same ``infrared embarrassment'' \cite{Callan:1977gz, Coleman:1985rnk, Polyakov:2004vp}. 
However, this problem arises due to the omission of interactions between current loops, and in the rigorous treatment using abelian duality (which takes interactions into account), the problem  does not exist. 
In this sense, our aim here is to present the analysis of the $U(1)$ lattice gauge theory in the way in which we envision a successful solution to the much harder problem of Yang--Mills theory.

The formulation of $U(1)$ lattice gauge theory we shall be studying is given by the Villain action
\begin{equation}
    S = \frac{1}{4e^2} \sum \lb \Delta_\mu A_\nu - \Delta_\nu A_\mu + 2\pi m_{\mu\nu}\rb^2
\end{equation}
$A_\mu$ is a vector field, $m_{\mu\nu}$ an antisymmetric tensor field.
The partition function integrates over all real values of $A_\mu$ and sums over all integer values of $m_{\mu\nu}$.
One can motivate the action by considering the weak coupling limit of the perhaps more familiar Wilson action given by
\begin{equation}
    S = \frac{1}{2e^2} \sum \lb 1 - \cos (\Delta_\mu A_\nu-\Delta_\nu A_\mu)\rb
\end{equation}
As $e\to0$, only configurations with $\Delta_\mu A_\nu-\Delta_\nu A_\mu$ near an integer multiple of $2\pi$ should be important, and for such configurations, we can approximate the cosine by a quadratic, leading to the Villain action.
However, we shall not be studying the weak coupling limit.
We are interested in the Villain model in its own right  at arbitrary  $e$ because of its resemblance to continuum gauge theory, and crucially, because of the clarity it brings to the role of magnetic monopoles.

The monopole degrees of freedom reside in the ``discrete magnetic fluxes'' $m_{\mu\nu}$; their worldlines are described by the current distribution
\begin{equation}
    k_\alpha = \tfrac12 \epsilon_{\alpha\lambda\mu\nu} \Delta_\lambda m_{\mu\nu}
\end{equation}
To isolate their contribution to the action, we employ the lattice analog of the Hodge decomposition on $m_{\mu\nu}$:
\begin{equation}
    m_{\mu\nu} = \Delta_\mu \xi_\nu - \Delta_\nu \xi_\mu + \epsilon_{\mu\nu\rho\sigma} \nabla_\rho \eta_\sigma
\end{equation}
Here $\eta_\mu$ is determined by the 4-vector Poisson equation
\begin{equation}
    G^{-1} \cdot \eta_\alpha = k_\alpha
\end{equation}
($G^{-1} = - \Delta_\mu\nabla_\mu$ is the lattice Laplacian), and with $\eta_\mu$ so-defined, the existence of $\xi_\mu$ is then guaranteed by the fact that $m_{\mu\nu}-\epsilon_{\mu\nu\rho\sigma}\nabla_\rho\eta_\sigma$ satisfies the Bianchi identity.
Substituting this decomposition for $m_{\mu\nu}$ in the action, we find
\begin{equation}
    \begin{split}
    S
    &= \frac{1}{4e^2} \sum \lb \Delta_\mu A_\nu - \Delta_\nu A_\mu +2\pi\Delta_\mu\xi_\nu- 2\pi\Delta_\nu\xi_\mu + \epsilon_{\mu\nu\rho\sigma} \nabla_\rho\eta_\sigma \rb^2 \\
    &= \frac{1}{4e^2} \sum \lb \Delta_\mu (A_\nu + 2\pi\xi_\nu) - \Delta_\nu (A_\mu + 2\pi\xi_\mu)\rb^2 + \frac{(2\pi)^2}{4e^2} \sum \epsilon_{\mu\nu\alpha\beta} \epsilon_{\mu\nu\rho\sigma} (\nabla_\alpha\eta_\beta) (\nabla_\rho\eta_\sigma) \\
    &= \frac{1}{4e^2} \sum \lb \Delta_\mu A'_\nu - \Delta_\nu A'_\mu \rb ^2 + \frac{(2\pi)^2}{2e^2} \sum (\nabla_\alpha \eta_\beta)^2 \\
    &= \frac{1}{4e^2} \sum \lb \Delta_\mu A'_\nu - \Delta_\nu A'_\mu \rb^2 + \frac{(2\pi)^2}{2e^2} \sum k_\beta \cdot G \cdot k_\beta
    \end{split}
\end{equation}
where we have put $A_\mu'=A_\mu+2\pi\xi_\mu$ and used the fact that $\eta_\mu$ is divergence-free. 
We see that the action has  decomposed into the sum of two decoupled terms, one the action of a free massless vector field, the other the action of an assembly of monopole worldlines with Coulombic interactions.
Changing the variable of integration from $A_\mu$ to $A_\mu'$ eliminates all dependence on $\xi_\mu$.
The sum over $m_{\mu\nu}$ can then be replaced by a sum over all integer current distributions $k_\mu$ satisfying the continuity condition $\Delta_\mu k_\mu =0$.%
\footnote{There is a degeneracy of $m_{\mu\nu}$ configurations that give rise to the same $k_\mu$ configuration, but the degeneracy is the same for all $k_\mu$ configurations.
Thus, the sum over $m_{\mu\nu}$ is just the sum over divergence-free $k_\mu$ times an overall factor which is of no consequence.}
Thus, the partition function of the model factorizes into the product of the partition function of free photons, $Z_{\gamma}$, and the classical statistical mechanics partition function of a Coulomb ``gas'' of monopole loops, 
\begin{equation}
    Z_{\text{mon}} = \sum_{\{k\}} \delta(\Delta_\mu k_\mu) \exp \biggl[-\frac{(2\pi)^2}{2e^2}\sum k_\mu(z) G(z-z') k_\mu(z') \biggr] 
    \label{interacting}
\end{equation}

\subsection{Monopole loop gas: its virtues and shortcomings}
Let us now examine the monopole loop gas partition function $Z_{\text{mon}}$ in more detail to understand the phase structure of the model.
We first explicitly separate out the $z=z'$ part of $G(z-z')$ and write 
\begin{equation}
    G(z-z') \equiv   \hat  G(z-z') + G_0 \delta(z-z')
\end{equation}
The term $G_0 \delta(z-z')$ corresponds to the self-energy of a loop segment and   
$\hat  G(z-z')$ corresponds to the interaction between distinct loop segments.

In what follows, we first ignore (unjustifiably, to see the harm done) the interaction part, and then we restore it in the subsequent section. 
We do this because it is entirely analogous to what is called the ``dilute instanton gas'' \cite{Callan:1977gz}. The logic goes as follows. Monopole loops are {\it finite action} configurations in $U(1)$ lattice gauge theory, similar to finite action instantons in non-abelian $SU(N)$ gauge theory. 
Furthermore, as shown in \S.\ref{sec:int},  the long distance interaction between small monopole current loops is of the form of magnetic dipole-dipole interaction  \eqref{magic0} and falls   as $1/|x|^4$.  This form  is identical to the long distance interaction between small instantons in non-abelian gauge theory \eqref{magic2}. In Yang-Mills theory, because of the fact that $1/|x|^4$ falls rather fast, 
the interaction between the instantons is entirely dropped and a non-interacting instanton gas is constructed.  Then, the integration over the size moduli of instanton leads to the famous infrared divergence problem \eqref{DIG}.  In $U(1)$ lattice gauge theory,  
we will see that the same problem arises, i.e., analysis gives IR-divergent result to vacuum energy density.  However, in the case of $U(1)$ gauge theory, we know that  $1/|x|^4$ arises as in the multipole expansion of current loops, and  microscopically, there is a $ \hat  G \sim 1/|x|^2$ interaction between current bits \eqref{interacting}.  This step is precisely what we do not know in Yang-Mills theory.    This provide a clear counter-example to the erroneous ``approximation'' in Yang-Mills theory.   In $U(1)$ gauge theory, in the regime 
where non-interacting loop gas gives divergent results, the interacting theory can be   dualized, and is free of infrared divergences.  In the  dual theory, it is easy to derive 
 mass gap and confinement. 
 
The partition function of the non-interacting loop gas can be written
\begin{equation}
    Z_{\text{mon}}^{\text{ideal}} = \sum_{\{k\}} \delta(\Delta_\mu k_\mu) \exp \biggl[ - \mu_0 \sum (k_\mu)^2 \biggr]
\end{equation}
where $\mu_0$ denotes the ``mass'' of the elementary monopole:
\begin{equation}
    \mu_0 \equiv \frac{(2\pi)^2}{2e^2} G_0
\end{equation}
While $k_\mu(z)$ can take any integer value, we shall focus only on the minimally charged loops for which $k_\mu(z)=0,\pm 1$, since the larger charged loops are exponentially suppressed in their Boltzmann weights. What we keep are the leading terms in the cluster expansion. 

Let us first focus on the single-monopole contribution to the partition function. 
A single monopole loop of length $L$ has action $S_0(L) = \mu_0 L$.  
This implies that the contribution of any one such loop is exponentially suppressed in its length as 
\begin{align}
    e^{-\mu_0 L}
\end{align}
However, the entropic factor $N(L)$ due to the multiplicity of all such loops is exponentially large in $L$.  
We have to take this entropic factor into account. 

To find this entropic factor, we must count the number of nonbacktracking loops of length $L$ that begin and end at a given site $x_0$.
We can estimate this number as follows.
Starting at the first site $x_0$, there are $2d$ choices ($d$ being the dimension of the lattice, 4 in our case) for selecting the second site along the loop.
At the second site, there are $2d-1$ choices for selecting the third site, since the loop is required to be nonbacktracking.
At each step after the first, there are $2d-1$ choices for selecting the next site, which is required to be distinct from the previous site.
Since there are $L$ sites on the loop, we have the following (over)estimate for the total number of nonbacktracking loops of length $L$ beginning and ending at a given site:
\begin{equation}
    N_{\rm up}(L) = (2d-1)^L = 7^L
\end{equation}
(We have ignored the fact that at the first step we actually had $2d$ choices rather than $2d-1$.)
This estimate is clearly an overestimate however, since it is actually counting paths that do not return to the starting point.
The reason, of course, is that we have assumed that each step, there were $2d-1$ choices for the next step.
This is certainly true for the first $L/2$ steps, but for every step after that, our choices are more constrained, because we have to start making our way back to the initial site.
We can therefore obtain an underestimate by counting the total number of nonbacktracking paths of length $L/2$ starting at $x_0$ but ending wherever, which is
\begin{equation}
    N_{\rm low} (L) = (2d-1)^{L/2} = 7^{L/2}
\end{equation}
The important point is that both the upper and lower bounds are exponentially growing in $L$, so we can assert that the true asymptotic growth in the number of closed loops of length $L$ is also exponentially growing in $L$,
\begin{equation}
    N(L) \approx e^{CL} \qquad (\half \ln 7  <  C < \ln 7)
\end{equation}
in $d=4$ dimensions.
Actually, the only thing that matters about the value of $C$ is that it is positive, which these arguments make clear is the case.

\subsection*{``Ideal'' monopole loop gas}
In analogy with the 1-instanton contribution to the partition function of Yang--Mills theory \cite{Callan:1977gz, Polyakov:1987ez, Coleman:1985rnk}, let us compute the 1-monopole contribution to the partition function. 
A single monopole loop can start and end at any point $x$ and can have any length $L>0$, with Boltzmann weight $e^{-\mu_0L}$ and density of states $e^{C L}$. 
Thus, the sum over all 1-monopole loop configurations is
\begin{equation}
    Z_{\text{1-mon}}  = \sum_{x} \sum_{L} e^{CL} e^{-\mu_0L}/L  \sim V \int_0^\infty dL \;  e^{(C-\mu_0)L}  /L
\label{1-monPF-mod }
\end{equation}
where $V$ denotes the spacetime volume and we have divided by $L$ since loops differing only in where they start and end along the loop should not be counted as distinct configurations. 
Since the full partition function of an ideal (noninteracting) gas of such monopole loops is simply the exponentiation of the 1-monopole contribution,
\begin{equation}
    Z_{\text{mon}}^{\text{ideal}} = \sum_{n=0}^{\infty} \frac{1}{n!} (Z_{\text{1-mon}})^n = \exp Z_{\text{1-mon}}
    \label{monPF}
\end{equation}
we see that the ``free energy density'' (adopting classical statistical mechanics language) is given by
\begin{equation}
    f_{\text{1-mon}}^{\text{ideal}} = -\ln Z_M^{\text{ideal}}/V = - \int_0^\infty \frac{dL}{L} \;  e^{(C-\mu_0)L}
    \label{free-energy}
\end{equation}
Remembering that $\mu_0\propto1/e^2$, we see that at sufficiently weak coupling that $\mu_0>C$, the integral over $L$ converges.
On the other hand, at sufficiently strong coupling that $\mu_0>C$, the integral over $L$ diverges!
This is exactly the same problem one encounters in the dilute instanton gas approximation in 4d Yang--Mills theory.


Physically, we understand what is happening.
The weak coupling regime corresponds to the Coulomb phase.
Here, monopole loops are small and rare since their energy dominates their entropy; they have no importance in the long distance physics.
The divergence in the integration over $L$ as the coupling surpasses the critical value $e_{\rm c}$ signals a phase transition, owing to the exchange of dominance of energy to entropy.
Once entropy wins, monopole loops of arbitrary size proliferate, and the phase becomes that of electric confinement. 

The divergence in the free-energy density is not in itself enough to deduce the occurrence of electric confinement, but it does draw a close analogy between the sum over monopole loop lengths, \eqref{monPF}, and the integral over instanton sizes in the dilute gas approximation of the Yang--Mills instanton calculus, \eqref{DIG}. 
In particular, a loop of length $L$, considering that it arises from a nonbacktracking random walk, will have a characteristic size  $\sim \sqrt{L}$. 
Thus, the integral over $L$ can also be viewed as an integration over the characteristic size of a monopole loop.
To repeat what was already mentioned in the introduction, which we have now explained more clearly:
 \begin{align}
{\rm monopole  \; loop \;  size:}  \;\; \Longrightarrow  &  \int dL \;  N(L)  =  {\rm divergent}  \cr 
 {\rm instanton \;  size \; moduli:} \;\; \Longrightarrow &   \int d \rho  \;   N(\rho) \;=
 {\rm divergent}
\label{mon-inst }
\end{align} 
In other words, both $SU(N)$ Yang--Mills theory and this strongly coupled lattice $U(1)$ gauge model suffer the same ``infrared embarrassment.''  
In fact, the embarrassment is even worse in the latter, because at sufficiently strong coupling the growth in the density of monopole loops is exponential in the length of the loop, while in the case of Yang--Mills theory, the growth in the density of instantons is power-like in the size parameter.

However, at this stage, we will do something different from our ancestors \cite{Callan:1977gz, Polyakov:1987ez, Coleman:1985rnk}.   Instead of backing-up, we will face our wrong-doings.  
In an instanton analysis, or statistical field theory analysis of vacuum structure, one should not be dismissive about the long-range interaction between topological configurations. 
In the case of instantons, small instantons at a large separation, interact via $1/|x|^4$ \eqref{magic2}. 
This is same as small monopole loops at large separation \eqref{magic0}. However, in the cases of monopole loops, we also know that the formula  \eqref{magic0} arises from the long distance limit of Coulombic current--current interactions:
\begin{align}
\sum_{z,z'} k_\mu(z) G(z-z') k_\mu(z')  \sim \int d^4z \;  d^4z' \;  K_\mu(z) G(z-z') K_\mu(z')
\end{align}
The crucial point is,   $1/|x|^4$ interaction between magnetic moments of small loops, there underlies the $G(x) \sim 1/|x|^2$ between current segments.  
This will allow us to solve the theory at {\it strong} coupling analytically and reliably by using abelian duality (or electric magnetic duality).  If we can figure out the counterpart of this step to decode the 4d instantons in Yang-Mills theory, we should have a decent  chance of solving the theory.

\subsection{Proliferation of interacting monopole loops: magnetic Higgs mechanism}

The divergence in the free energy density was the result of blatantly disregarding interactions between monopole loops.
It is now time to restore this interaction:
\begin{equation}
    Z_{\text{mon}} = \sum_{\{k\}} \delta(\Delta_\nu k_\nu) \exp \biggl[ - \frac{(2\pi)^2}{2e^2} \sum k_\nu \cdot \hat G \cdot k_\nu - \mu_0 \sum (k_\nu)^2 \biggr] 
\end{equation}
This can also be written as a Gaussian integral over a vector field $B_\mu$, which should be interpreted as the dual potential of the gauge field:
\begin{equation}
    Z_{\text{mon}} = \int \lb dB \rb \; \exp \biggl[ - \frac{e^2}{2(2\pi)^2} \sum B_\mu \cdot \hat G^{-1} \cdot B_\mu \biggr] Z_{M}\lb B\rb
    \label{eq:worldline}
\end{equation}
where 
\begin{equation}
    Z_{M}\lb B\rb = \sum_{\{k\}} \delta( \Delta_\nu k_\nu) \exp \biggl[i \sum B_{\nu} k_{\nu} - \mu_0 \sum (k_\nu)^2 \biggr] 
    \label{eq:mpartfn}
\end{equation}
(Note that $Z_{M}\lb B=0\rb$ is just what we earlier called $Z_{\text{1-mon}}$, the 1-monopole partition function.)

As we will now show, the partition function of the model as represented in \eqref{eq:worldline} and \eqref{eq:mpartfn} can be formally identified with the worldline representation of a lattice scalar QED in which the scalar field is magnetically coupled to the gauge field. 
We then argue that sufficiently strong coupling in the Villain model we started with corresponds to the Higgs phase of this scalar QED. 
In other words, the strong coupling phase of the Villain gauge theory entails the condensation of monopoles, and this is enough to imply electric confinement.

\begin{figure}[t]
\begin{center}
 \includegraphics[angle=0, width=0.8\textwidth]{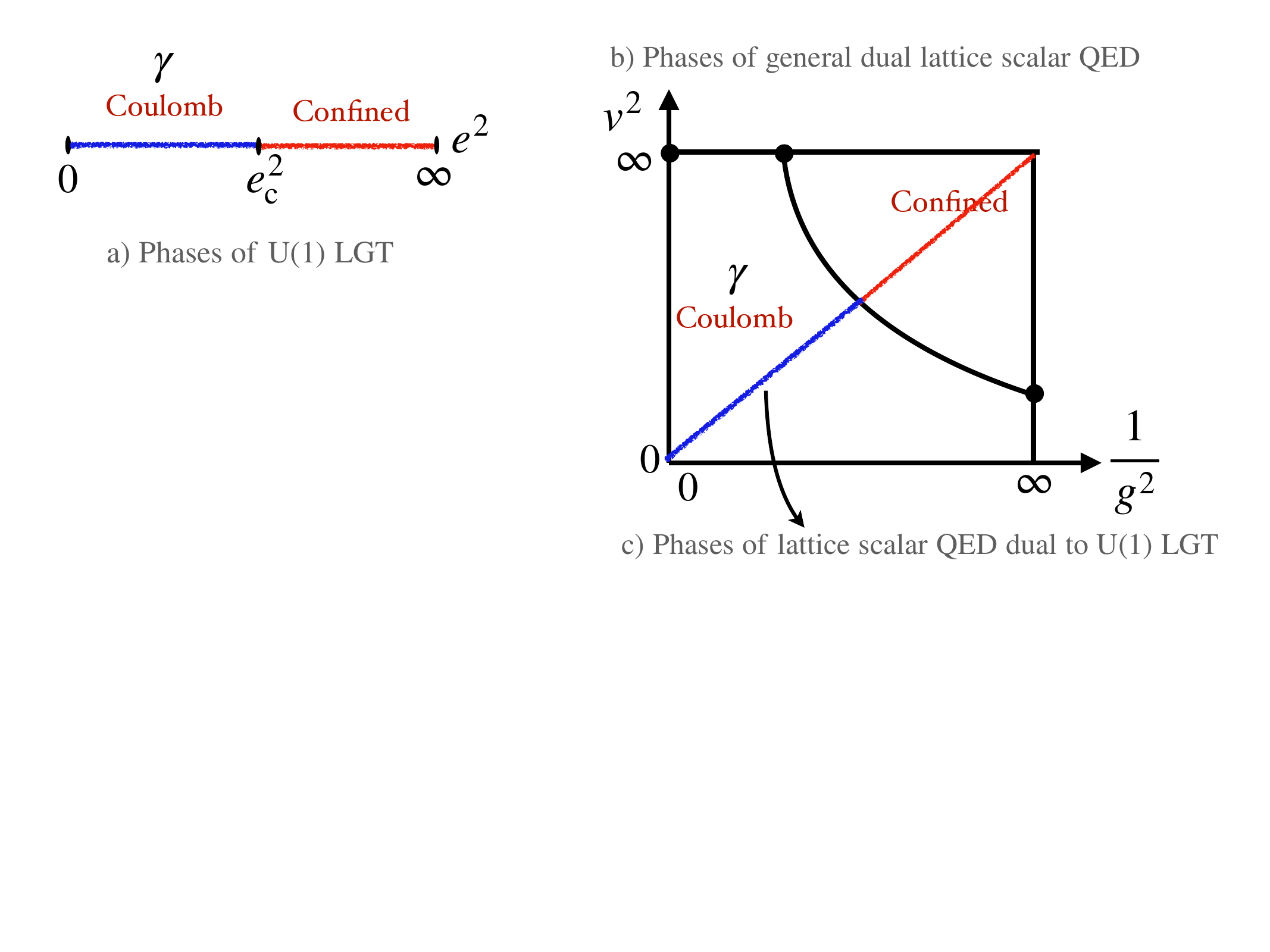}
\vspace{-4cm}
\caption{a) $U(1)$ lattice gauge  in 4d theory has two phases. A  strong bare coupling confined phase  and a weak bare coupling Coulomb phase.  b) The (magnetic) lattice scalar QED model in 4d also has two phases at fixed but finite $\lambda$,   confined and Coulomb.  c) The dual of $U(1)$ LGT corresponds to the line shown in the figure. The parameters  $g^2$ and $v^2$  along the dual line are dictated by $e^2$ of the original $U(1)$  lattice gauge theory.      }
\label{duality}
\end{center}
\end{figure}

So let us for the moment leave aside the Villain model and consider now lattice scalar QED.
We shall write the partition function as
\begin{equation}
    Z_{\text{scalar QED}} = \int \lb dB \rb \; \exp\biggl[-\frac{1}{2g^2} \sum B_\nu \cdot \hat G^{-1} \cdot B_\nu \biggr] Z_\phi\lb B \rb
\end{equation}
where
\begin{equation}
    Z_\phi\lb B \rb = \int \lb d\phi \rb \lb d\phi^* \rb \; \exp\biggl[ \sum_{z,\nu} \phi_{z-\nu} e^{iB_{z,\nu}} \phi^*_z + \sum_{z,\nu} \phi^*_{z-\nu} e^{-iB_{z,\nu}} \phi_z - \sum_z V(|\phi_z|) \biggr]
\end{equation}
Here, $\phi$ is a scalar field and $B_\nu$ is a vector field that we interpret as the dual potential of the gauge field, which means that $\phi$ is magnetically charged, and $g$ is the magnetic coupling constant.
For the scalar potential $V$, we can take something like 
\begin{equation}
    V(r) = \lambda (r^2 - v^2)^2
\end{equation}

Our goal is to integrate out the matter field and obtain an expression in terms of  Wilson loops of the background gauge field, i.e., convert the matter sector into a sum over loops. 
Performing a Taylor series expansion  for the hopping terms, we find:
\begin{multline}
    \exp \biggl( \sum_{z,\nu} \phi_{z-\nu} e^{iB_{z,\nu}} \phi^*_z + \sum_{z,\nu} \phi^*_{z-\nu} e^{-iB_{z,\nu}} \phi_z \biggr)
    \\
    = 
    \sum_{\smash{\{p_{z,\nu}\}}} 
    \sum_{\smash{\{q_{z,\nu}\}}} 
    \prod_{z,\nu} 
    \frac{e^{i B_{z,\nu} (p_{z,\nu} - q_{z,\nu})}}{p_{z,\nu}!\, q_{z,\nu}!} 
    (r_z)^{p_{z+\nu,\nu} + q_{z,\nu} + p_{z,\nu} + q_{z+\nu,\nu}} 
    e^{i \theta_z (p_{z+\nu,\nu}+q_{z,\nu} - p_{z,\nu} - q_{z+\nu,\nu}) } 
    \label{powerexp}
\end{multline}
where we have introduced the radial and angular parts of the scalar field, $\phi=re^{i\theta}$, and the sums over $p_{z,\nu},  q_{z,\nu} \in [0, \infty)$, i.e. they  run through the non-negative integers. 

The path integral measure is $\lb d \phi\rb\lb d \phi^*\rb =\prod_{z} r_z dr_z d\theta_z$. 
The integration over the  angular parts can be performed exactly,  and for the radial parts, we will use the saddle point method. 
 The integration over $ \theta_z$ yields a constraint, a product of Kronecker deltas of the form 
 $\delta\bm( \sum_{\nu}  (p_{z+\nu,\nu}+q_{z,\nu} - p_{z,\nu} - q_{z+\nu,\nu})\bm)$.  
This constraint is nothing but current conservation, and helps us identify monopole currents 
in terms of $p_{z,\nu}$,   and $q_{z,\nu}$. 
To see this, it is useful to define new variables
\begin{equation}
    k_{z,\nu} \equiv p_{z,\nu} - q_{z,\nu}, \quad 2 \ell_{z,\nu} + |k_{z,\nu}| \equiv p_{z,\nu} + q_{z,\nu}, \qquad  k_{z,\nu} \in (-\infty,\infty), \;\;  \ell_{z,\nu}
    \in [0,\infty),
\end{equation}
Then we can rewrite \eqref{powerexp} as 
\begin{equation}
    \sum_{\{k_{z,\nu}\}} 
    \sum_{\{\ell_{z,\nu}\}}
    \prod_{z,\nu} 
    \frac{e^{i B_{z,\nu} k_{z,\nu}}}{\ell_{z,\nu}! (|k_{z,\nu}| + \ell_{z,\nu})!} 
    (r_z)^{ 2 \ell_{z+\nu,\nu} + |k_{z+\nu,\nu}| + 2 \ell_{z,\nu} + |k_{z,\nu}|} 
    e^{i \theta_z (k_{z+\nu,\nu} - k_{z,\nu} ) }
\end{equation}
Integrating out $\theta_z$, the last exponential factor gets replaced by the delta functional $\delta( \fdif_\nu k_{\nu} )$.
The Higgs partition function then becomes
\begin{equation}
    Z_\phi\lb B\rb = \sum_{\{k\}} \delta (\Delta_\nu k_\nu) \exp\biggl(i \sum_{z,\nu} B_{z,\nu} k_{z,\nu} \biggr) \sum_{\{\ell\}} \biggl[ \prod_z I(f_z) \prod_\nu \frac{1}{\ell_{z,\nu}! (|k_{z,\nu}| + \ell_{z,\nu})!} \biggr]
    \label{eq:hpartfn}
\end{equation}
where we have put
\begin{equation}
    I(f) \equiv \int_0^\infty e^{-V(r)} r^{f+1} dr, \qquad    f_z \equiv \sum_\nu (2 \ell_{z+\nu,\nu} + |k_{z+\nu,\nu}| + 2 \ell_{z,\nu} + |k_{z,\nu}|)
    \label{radial}
\end{equation}

We  can  perform the radial integral using the saddle point analysis.  
Deep in the Higgs phase, the critical point $r = v$ dominates the integral and yields
\begin{equation}
    \prod_{z}  (v)^{ f_z}   =    \prod_{z}  (v)^{  \sum_{\nu} (4 \ell_{z, \nu} + 2 |k_{z+\nu,\nu}| )} \;.
   \label{aux}
\end{equation}  
Therefore, the sum over loops in $Z_\phi$, which can be viewed as the weight factor,  can be  expressed   as: 
\begin{equation}
    W\lb k\rb   =     \sum_{ \{ \ell \} }    
    \left( \prod_{z,  \nu}  \frac{1 } { \ell_{z, \nu}!  ( |k_{z, \nu}| + \ell_{z, \nu})!} \right) 
  \left( \prod_{z \nu} (v^2)^{   (2 \ell_{z, \nu} +  |k_{z+\nu,\nu}| ) }  \right)   
   \label{weight2}
\end{equation} 
The sum over  $\ell_{z, \nu}$ can be performed explicitly and gives the modified Bessel function for each $k_{z, \nu}$:
        \begin{align} 
 I_{k_{z, \nu}}(2 v^2)   =  \sum_{ \ell_{z, \nu} }    
  \frac{1 } { \ell_{z, \nu}!  ( |k_{z, \nu}| + \ell_{z, \nu})!} 
 (v^2)^{   (2 \ell_{z, \nu} +  |k_{z+\nu,\nu}| ) }    , \qquad  I_{|k_{z, \nu}|} = I_{k_{z, \nu}}  \quad  \forall l_{n, \nu} \in \Z
   \label{weight3}
 \end{align} 
 Hence, the weight factor  takes a simple  form:
        \begin{align} 
 & W[k]   =    
    \prod_{z,  \nu}    I_{k_{z, \nu}}(2v^2) 
   \label{weight4}
 \end{align} 
which lead to  the partition function of the magnetic matter sector in a background gauge field: 
\begin{equation}
    Z_\phi \lb B \rb   = \sum_{ \{ k  \}}  \delta ( \Delta_\nu k_\nu) \exp\biggl[i\sum B_\nu k_\nu \biggr]  \prod_{z,  \nu}   I_{k_{z, \nu}}(2v^2)
\end{equation}
Using large $ v$ asymptotics  of the Bessel function,    in the deep magnetic Higgs phase, 
         \begin{align} 
 I_{k_{z, \nu}}(2 v^2)   \sim 
 \frac{1}{\sqrt{4 \pi v^2}}  e^{2 v^2}    e^{- {(k_{z,\nu} )^2}/{2 v^2}} ,   \qquad v \rightarrow \infty  
    \label{weight5}
 \end{align}  
 $Z_{\phi}\lb B\rb$ reduces to the same expression as the worldline representation of  $U(1)$ lattice gauge theory in terms of  monopole loops in \eqref{eq:mpartfn}
\begin{equation}
    Z_\phi \lb B \rb = \sum_{\{k\}} \delta(\Delta_\nu k_\nu) \exp \biggl[i \sum B_\nu k_\nu - \frac{1}{2v^2} \sum (k_\nu)^2 \biggr] 
    \label{eq:weight6}
\end{equation} 
with  identification: 
\begin{equation}
  \frac{1}{2v^2}=   \mu_0 \equiv \frac{(2\pi)^2}{2e^2} G_0  
\end{equation}
Comparing with \eqref{eq:mpartfn}, we see that deep in the (magnetic) Higgs regime, $v \to \infty$, we have 
\begin{align}
    Z_\phi\lb B \rb \approx Z_{M}\lb B\rb
\end{align}
 if the modulus $v$ is identified with the electric coupling $e$ times some overall constant.
The proliferation of magnetic loops therefore corresponds to the condensation of a magnetically charged scalar field.


Note that in the original $U(1)$ lattice gauge theory, we only have one coupling, $e^2$. 
A generic scalar QED model  has three independent couplings, 
$g^2$, $v^2$ and $\lambda$.  However, our dual formulation is not the generic 
 model, and the couplings are not independent.   In our dual formulation,  these  couplings are all determined in terms of one coupling of the lattice gauge theory, $e^2$.  In this sense, the phase diagram of $U(1)$ lattice gauge theory map to a line in the phase diagram of a general scalar QED as shown in   Fig.\ref{duality}.

Also note an interesting, but somewhat well-known lattice artifact \cite{Borgs, Borgs:1985ik}. A finite vev  $v^2$ for the lattice Higgs scalar does not necessarily imply condensation. In fact,  at fixed and large   $v^2$, if one takes $g^2$  large, scalar QED exhibits Coulomb phase, not condensation \cite{Borgs:1985ik}.  At large   $v^2$, with  sufficiently small  $g^2$, we observe a condensed phase. 
Of course, the  $e^2 \ll e_c^2$ regime  of $U(1)$ lattice gauge theory  corresponds to the limit where 
$v^2 \sim \frac{1}{g^2} \ll 1 $ and the  Coulomb phase is realized, while 
the strong coupling  $e^2  \gg e_c^2 $ limit correspond to the regime 
$v^2 \sim \frac{1}{g^2 } \gg 1  $ are simultaneously large, and the (magnetic) condensation phase is realized.

\subsection{Interactions between small current loops}
\label{sec:int}
 As seen in the world-line formalism, in the Coulomb phase, the magnetic current loop  remain small and do not proliferate. This is a somewhat boring phase dynamically.  In strong coupling, there exist both small and large loops, and larger ones are  entropically favored. 
Below, we calculate the interaction between small magnetic current loops at large separations, which has the same form as the interaction between instantons.   
 Consider two small magnetic current loops (these should be thought in the context of magnetostatics in 4+1 dimensions.)
 \begin{align}
 K^{(i)}_\mu(x) = g  \int d \tau \frac{dx_\mu^{(i)}}{d \tau} \delta^4\bm(x- x^{(i)}(\tau)\bm),  \quad i=1,2 
 \label{mcur}
\end{align}
 well-separated from each other.  Here, $g = \frac{2 \pi}{e}$. 
 We can write the interaction between two currents loops as: 
 
 We can write the interaction between them as 
  \begin{align} 
V_{\rm loop-int} = \int d^4x   \;  K^{1}_\mu[x] B_{\mu}(x)  
=  \int d^4x d^4y   \;  K^{1}_\mu[x] G(x-y)  K^{2}_\mu[y] 
\label{loopint}
\end{align}
where $B_{\mu}(x)$ is the vector potential induced by the $K^{2}_\mu[y]$ at $x$: 
  \begin{align} 
 B_{\mu}(x)  =  \int  d^4y  \;  G(x-y) K^{2}_\mu[y] 
\label{duala}
\end{align}
Substituting  \eqref{mcur} into \eqref{loopint},  we can express the interaction between distant small loops as double-integral along the loop worldlines: 
  \begin{align} 
V_{\rm loop-int} =  g^2 \int d\tau d\sigma  \;   \dot x_\mu^1(\tau)  \dot x_\mu^2(\sigma) \;   G( x^1(\tau) -x^2(\sigma)   ) 
\end{align}
Let us decompose the coordinates of the loop   $x_\mu^i(\tau)$ into a center position  $x_\mu^i$  and  reparametrization of the loop with respect to that central point $ \xi_{\mu}^i(\tau)$: 
\begin{align}
 x_\mu^i(\tau) \equiv  x_\mu^i + \xi_{\mu}^i(\tau) 
\end{align} 
 We assume
$|x^1 - x^2| \gg |\xi_{\mu}^i(\tau)| $, i.e. the separation between loops is  much larger than the characteristic sizes of the loops.  Hence, we can write interactions  between the two loop as 
  \begin{align} 
V_{\rm loop-int}& =  g^2\int d\tau d\sigma   \dot \xi_\mu^1(\tau)  \dot \xi_\mu^2(\sigma)  G(x^1 - x^2 +   \xi^1(\tau) -\xi^2(\sigma)   )   \cr
 &=  g^2 \left(  \int d\tau   \dot \xi_\mu^1(\tau)    \xi^1_{\alpha}(\tau)  \right)  \left( \int   d\sigma  \dot \xi_\mu^2(\sigma) \xi_\beta^2(\sigma)   \right)     \partial_{\alpha} \partial_\beta G(x^1 - x^2)  \cr
& =   g^2 \left(  \int d\tau   \dot \xi_{[\mu }^1(\tau)    \xi^1_{\alpha ]}(\tau)  \right)  \left( \int   d\sigma  \dot \xi_{[\mu}^2(\sigma) \xi_{\beta]}^2(\sigma)   \right)     \partial_{\alpha} \partial_\beta G(x^1 - x^2)  \cr 
&= g^2 M^1_{\mu \alpha} M^2_{\mu \beta}   \partial_{\alpha} \partial_\beta G(x^1 - x^2) \cr 
& =  \frac{g^2}{2 \pi^2  |x|^4}   \left( 4 ( M^{1}_{{\mu \alpha}}  \hat x_\alpha )(   M^{2}_{\mu \beta}    \hat x_\beta ) -   M^{1}_{{\mu \alpha}}  M^{2}_{{\mu \alpha}}  \right)
\label{loopint2}
\end{align}
where in the third line, we used the fact that 
 $\int d\tau   \dot \xi_{(\mu }^1(\tau)    \xi^1_{\alpha )}(\tau) =  \int d\tau \frac{\partial}{ \partial \tau}  ( \xi_{(\mu }^1(\tau)    \xi^1_{\alpha )}(\tau) )= 0$.  As expected, this is the interaction between two magnetic dipoles sourced by magnetic currents in magnetostatics in 4+1 dimensions. 

 The Coulomb phase with dilute dipoles   $M_{\mu \beta}$ seems somewhat boring.  In particular, $\frac{1}{|x|^4}$ type interaction between the dipoles cannot destroy the Coulomb phase at weak coupling.  However, the crucial point is following. The very same objects, that are rather 
 unimportant in weak coupling,   once their populations and size grows due to entropic factor, becomes the microscopic mechanism of confinement in the magnetic Higgs phase (corresponding to strong coupling on lattice)!    In this regime,  the basic classical interaction between current bits is $G(x) \sim \frac{1}{|x|^2}$. 
It would be a major mistake to dismiss the underlying interaction, and declare them irrelevant for confinement  based on the  long range $\frac{1}{|x|^4}$    interaction between small loops at large separations.

\section{Yang-Mills instantons versus self-dual monopole loops}
Instantons are saddle points of path integrals that are localized in Euclidean space-time \cite{Belavin:1975fg}. In a certain sense, they are supposed to be viewed as point-like objects in space-time.  Monopoles in 3+1 gauge theory are particles, and they are described by their world-lines (in our context as closed loops)  in space-time. As such, the dimensionalities of these objects do not match. However, in certain asymptotically free quantum field theories such as $\mathbb CP^{N-1}$ in 2d, and asymptotically free gauge theories  without elementary scalars in 4d (such as Yang-Mills, QCD, ${\cal N}=1$ supersymmetric gauge theory), the instanton size $\rho$  is part of its moduli space, i.e, the action of the configuration does not depend on its size. In Yang-Mills theory, the moduli space is, to repeat \eqref{moduli} 
\begin{align}
  {\cal M} =  \mathbb R^4 \times \mathbb R^+ \times   \frac{SU(N)}{S\bm( U(N-2) \times U(1)\bm)}
  \label{moduli2}
\end{align}
The crucial point is that in the path integral, we are integrating over all sizes of instantons. In this sense, at least in this class of theories, instantons should not be viewed as events localized in space-time, rather a configuration with characteristic size 
$\rho $ where $\rho$ varies in between $[0, \infty]$.  

Below, we derive the instanton operator on $\R^4$ in the leading semiclassical approximation, as by product obtaining the leading {\it classical} interactions between instantons and anti-instantons for gauge group $SU(N)$. For gauge group $SU(2)$, this was calculated in \cite{Callan:1977gz}.  As we show in detail, the interaction is  crucially {\it long ranged},
the one between two  
{\it non-abelian  magnetic dipoles}, respectively, anti-selfdual dipole moment for instanton, and  selfdual dipole moment for anti-instanton.  We will show that this dipole moment can be viewed as if it is emanating from a network of self-dual magnetic current loops (by this, we mean combination of currents that induces self-dual fields). 


\subsection{Classical interactions between instantons}

Our strategy for computing the leading large distance classical interactions between instantons and themselves or anti-instantons follows that of Ref.~\cite{Vainshtein:1981wh}.%
\footnote{
Our presentation simplifies that of \cite{Vainshtein:1981wh} in that we work entirely in Euclidean space, whereas they begin in Minkowski space and invoke the LSZ theorem.
}

The first step is to obtain an expression for the effective instanton operator.
By an effective instanton operator, we mean the following.
Suppose field configurations of characteristic wavelength less than some length $\rho_0$ have been integrated out.
Then in particular, instantons of radius less than $\rho_0$ have been integrated out.
Their imprint on the effective action for modes of wavelengths $> \rho_0$ appears as a term of the form
\begin{equation}
    \Delta S_{\rm inst} = \int d^4 x \int_{\hat{\mathcal M}_{\rho_0}} d^{4N-4} w \; K_w e^{-S_0} \mathcal I_w(x)
\end{equation}
where $\hat{\mathcal M}_{\rho_0}$ is the subregion of the centered instanton moduli space $\hat{\mathcal{M}}$ that parametrizes instanton configurations of radius $<\rho_0$, and $K_w$ is a factor coming from the functional determinant of the fluctuation operator in the instanton background. 
The local operator $\mathcal I_w$, which is parametrized by the internal collective coordinates $w$, is the effective instanton operator. 
It is defined by the property that for any multilocal operator $\mathcal{O}(x_1,\ldots,x_n)$  (any product of gauge fields and derivatives at the points $x_1,\ldots,x_n$), the expectation value of $\mathcal I_w(X) \mathcal{O}(x_1,\ldots,x_n)$ computed in the effective theory gives the contribution of a single instanton with internal parameters $w$ at $X$ to the expectation value of $\mathcal{O}(x_1,\ldots,x_n)$.

Once we have the effective instanton operators $\mathcal I$, as well as the similarly defined anti-instanton operators $\overline{\mathcal I}$, we can extract the interaction $S_{\rm IA}$ or $S_{\rm II}$ between an instanton with an anti-instanton or with another instanton via
\begin{equation}
    e^{-S_{\rm IA}} = \langle \mathcal I(X) \overline{\mathcal I}(Y) \rangle_{\rm eff}\qquad\text{or}\qquad
    e^{-S_{\rm II}} = \langle \mathcal I(X) \mathcal I(Y) \rangle_{\rm eff}
\end{equation}

Of course, we cannot find $\mathcal I_w$ exactly, but if $\rho_0$ is small, then we are justified in using the leading semiclassical approximation to compute the one-instanton contribution of $\mathcal{O}(x_1,\ldots,x_n)$, which simply amounts to substituting  the instanton solution $\mathcal A_{\mu} (x-X;w)$ for $A_{\mu}(x)$ everywhere in $\mathcal O(x_1,\ldots,x_n)$.
We then need to match this with the tree-level expectation value of $\mathcal I_w(X)\mathcal{O}(x_1,\ldots,x_n)$ to leading order in large $|x_i - X|$. 
Thus, in leading semiclassical order, the instanton operator is characterized by the condition
\begin{equation}
    \Bigl\langle \mathcal I_w(X) \mathcal{O}(x_1,\ldots,x_n) \Bigr\rangle_{\text{tree}} \approx \mathcal{O}(x_1,\ldots,x_n)\Bigr|_{A_\mu(x) = \mathcal{A}_\mu(x-X;w)} 
\end{equation}
where by ``$\approx$'' we mean that all that is required is asymptotic equality for large $|x_i-X|$. 
In fact, we can simply take the multilocal operator to be a simple product of gauge fields, so that the condition simply reads
\begin{equation}
    \Bigl\langle \mathcal I_w(X) \prod_{i=1}^n A^{a_i}_{\mu_i}(x_i) \Bigr\rangle_{\text{tree}} \approx \prod_{i=1}^n \mathcal A^{a_i}_{\mu_i}(x_i-X;w)
    \label{matching}
\end{equation}

We will now show that in leading semiclassical order, the instanton operator is given by%
\footnote{
We use the 4d Euclidean $\sigma$ matrices $\sigma_\mu = (1,i\vec \tau)$, where $\vec\tau$ is 3-vector of the standard Pauli matrices.
We also use $\sigma_{\mu\nu} = \tfrac14 (\sigma_\mu \sigma_\nu^\dag - \sigma_\nu\sigma_\mu^\dag)$ which is self-dual and $\bar\sigma_{\mu\nu}=\tfrac14 (\sigma_\mu^\dag\sigma_\nu - \sigma_\nu^\dag\sigma_\mu)$ which is anti-self-dual.
}
\begin{equation}
    \mathcal I = {:}\exp (-4\pi^2 e^{-1} \tr w \bar\sigma_{\mu\nu} w^\dag F_{\mu\nu}){:}
    \label{inst-op}
\end{equation}
We simply need to verify that it satisfies the matching condition \eqref{matching}.%
\footnote{
The deductive route to the formula proceeds by essentially running our discussion here in reverse.
}
As should be clear, it is really only necessary to work it out for the case of a two-point correlator of $\mathcal I$ and a single gauge field.
We start with the tree-level calculation in the effective theory: 
\begin{equation}
    \begin{split}
    \langle A_\mu^a(x) \mathcal I(X) \rangle
    &= -4\pi^2e^{-1} \langle A_\mu^a (x) F^b_{\nu\alpha}(X) \rangle \tr w \bar\sigma_{\nu\alpha} w^\dag T^b \\
    &\approx 8\pi^2e^{-1} \biggl\langle A_\mu^a(x) \frac{\del}{\del X_\nu} A_\alpha^b(X) \biggr\rangle \tr w \bar\sigma_{\nu\alpha} w^\dag T^b \\
    &= -8\pi^2e^{-1} \frac{\del}{\del X_\nu} \biggl( \frac{\delta^{ab}\delta_{\mu\alpha}}{4\pi^2|x-X|^2} \biggr) \tr w \bar\sigma_{\nu\alpha} w^\dag T^b \\
    &= -e^{-1} \frac{4 (x-X)_\nu}{|x-X|^4} \tr w \bar\sigma_{\nu\mu} w^\dag T^a
    \end{split}
\end{equation}
Comparing this with the formula for the instanton potential,
\begin{equation}
    A^a_\mu (x) = \mathcal A^a_\mu (x;w) \equiv e^{-1} \frac{4 (x-X)_\nu}{|x-X|^2(|x-X|^2+\rho^2)} \tr w \bar \sigma_{\mu\nu} w^\dag T^a
\end{equation}
we see that preceding two formulas are asymptotically equal at large $|x-X|$.
This confirms the correctness of \eqref{inst-op}.

An essentially identical calculation shows that
\begin{equation}
    \overline{\mathcal I} = {:} \exp (-4\pi^2 e^{-1} \tr v \sigma_{\mu\nu} v^\dag F_{\mu\nu} ) {:}
\end{equation}
is the correct formula for the anti-instanton operator to the order we are working.
It is simply the result of replacing $\bar\sigma_{\mu\nu}$ by $\sigma_{\mu\nu}$ in \eqref{inst-op} (as well as the internal instanton parameters $w$ by internal anti-instanton parameters $v$). 

We now turn to computing the classical instanton interactions.
Using the defining property of the instanton operator, we find
\begin{equation}
    \begin{split}
    S_{\rm II}(X-Y) 
    &= - \ln \langle \mathcal I_w(X) \mathcal I_{w'}(Y) \rangle \\
    &= - \ln \mathcal I_{w'}(Y)|_{A_\mu(x) = \mathcal A_\mu(x-X;w)} \\
    &=  4 \pi^2 e^{-1} \tr \lb w' \bar\sigma_{\mu\nu} w'^\dag \mathcal F_{\mu\nu} (X-Y;w) \rb
    \end{split}
\end{equation}
Vanishing of the interaction $S_{\rm II}$ between two instantons follows from the self-duality of $\mathcal F_{\mu\nu}$ and the anti-self-duality of $\bar\sigma_{\mu\nu}$. 
This, of course, was to be anticipated based on the known structure of the multi-instanton moduli space. 
The instanton--anti-instanton interaction $S_{\rm IA}$ is certainly nonzero, however.
By a similar calculation, we find
\begin{equation}
    S_{\rm IA}(X-Y) = 4 \pi^2 e^{-1} \tr \lb v \sigma_{\mu\nu} v^\dag \mathcal F_{\mu\nu} (X-Y;w) \rb
\end{equation}
To write it more explicitly, it is useful to note that the large $|x-X|$ behavior of the instanton field strength can be written
\begin{align}
    \mathcal F^a_{\mu\nu}(R;w) \approx& -16 \pi^2 e^{-1} \del_\alpha \del_{[\mu} G(R) (\tr w \bar\sigma_{\nu]\alpha} w^\dag T^a)  \cr
    = &- \frac{2}{\pi^2 }  \frac{1}{R^4} ( 2 {\hat R}_{\mu}  (\bar M_{\nu \alpha}^a {\hat R}_{\alpha}) - 2 {\hat R}_{\nu}  (\bar M_{\mu \alpha}^a {\hat R}_{\alpha}) - \bar M_{\nu \mu}^a)
\end{align}
where we have put $R=X-Y$. 
Then the interaction $S_{\rm IA}$ can be written as
\begin{align}
    S_{\rm IA}
   & = -(-8\pi^2 e^{-1} \tr v\sigma_{\mu\nu} v^\dag T^a) \del_\alpha \del_\mu G(R) (-8\pi^2 e^{-1} \tr w \bar\sigma_{\nu\alpha} w^\dag T^a)  \cr
  &  =  \frac{1}{2 \pi^2 e^2 R^4} ( 4 (M_{\nu \mu}^a \hat R_{\mu})( \bar M_{\nu \alpha}^a \hat R_{\alpha})  - M_{\nu \mu}^a  \bar M_{\nu \mu}^a )
\end{align}
We recognize here the form of the interaction energy between distant magnetic dipoles in a world of four spatial dimensions.
We associate the instanton with the \emph{anti-self-dual} magnetic dipole moment 
\begin{equation}
    \bar M_{\mu\nu}^a(w) = -8\pi^2 \tr w \bar\sigma_{\mu\nu} w^\dag T^a
\end{equation}
the anti-instanton with the \emph{self-dual} magnetic dipole moment
\begin{equation}
    M_{\mu\nu}^a(v) = - 8\pi^2 \tr v \sigma_{\mu\nu} v^\dag T^a
\end{equation}









\subsection{Self-dual fields in Abelian gauge theory}

The field of an instanton essentially becomes Abelian at asymptotically large distances from the center, at least in the sense that the decay of the commutator terms in the field-strength is more rapid than that of the derivative terms.\footnote{Here, we do not mean  that the non-abelian $SU(N)$ field at large distance reduces to $U(1)^{N-1}$, the  maximal abelian subgroup.  Rather, we mean that the field strength at leading order at  large distance can be written as a combination of  $(N^2 -1)$ $U(1)$ fields. i.e. $[A_{\mu}, A_{\nu}]$  term is suppressed compared to $\partial_{\mu}A_{\nu} -  \partial_{\nu}A_{\mu}$ term for fixed size instantons. The former scale as $\frac{\rho^2}{|x|^6}$ while the latter is $\frac{1}{|x|^4}$.  }
In view of this, it is natural to ask if it is possible to reproduce the large-distance form of Yang--Mills instantons in an Abelian gauge theory.
It is well-known that the vacuum Maxwell equations have no self-dual or anti-self-dual solutions that vanish at infinity. 
However, such solutions do exist if we introduce ``self-dual'' or ``anti-self-dual'' sources. 
        
The Euclidean Maxwell equations with both electric and magnetic sources read
\begin{equation}
    -\del_\mu F_{\mu\nu} = J_\nu,\quad -\del_\mu \tilde F_{\mu\nu} = K_\nu 
    \label{MaxwellEM}
\end{equation}
By a self-dual or anti-self-dual source, we mean that the ``electric'' and magnetic sources are respectively equal or opposite:%
\footnote{
The terminology is perhaps an abuse of language.
Strictly speaking, self-- and anti-self-duality is a notion that applies to antisymmetric two-index tensors.
All that is meant is that the fields produced by such sources are respectively self-dual or anti-self-dual. The reason for quotation marks around electric will be explained in the final subsection. 
}
\begin{equation}
    J_\nu = \pm K_\nu
    \label{MaxSD}
\end{equation}
For such self-dual or anti-self-dual sources, the field strength is obviously self-dual or anti-self-dual.
Here, we shall be interested in the behavior of the field very far from the sources.

To do this, we exploit the linearity of Maxwell's equations and consider separately the ``purely electric'' source problem 
\begin{equation}
    -\del_\mu F'_{\mu\nu} = J_\nu, \quad -\del_\mu \tilde F'_{\mu\nu} = 0
\end{equation}
and the ``purely magnetic'' source problem
\begin{equation}
    -\del_\mu F''_{\mu\nu} = 0, \quad -\del_\mu \tilde F''_{\mu\nu} = \pm J_\nu
\end{equation}
obtaining the solution to the self-dual or anti-self-dual problem by superposition: $F_{\mu\nu} = F'_{\mu\nu} + F''_{\mu\nu}$. 

The analysis is essentially a generalization of elementary magnetostatics to a world of four spatial dimensions.
Let us first consider the case of a purely electric source.
Since $\del_\mu \tilde F'_{\mu\nu}=0$, we are permitted to express the field-strength in terms of a potential:
\begin{equation}
    F'_{\mu\nu} = \del_\mu A_\nu - \del_\nu A_\mu
\end{equation}
Imposing the gauge condition $\del_\mu A_\mu = 0$, the Maxwell equations reduce to
\begin{equation}
    -\del^2 A_\nu = J_\nu
\end{equation}
and the solution is
\begin{equation}
    A_\nu(x) = \int G(x-y) J_\nu(y) \; d^4y
\end{equation}
where $G$ is the Green function of the ordinary Laplacian $-\del^2$. 
From this, one can find that the leading behavior of the field-strength asymptotically far from the source is
\begin{equation}
    F'_{\mu\nu}(x) = 2 \del_\lambda \del_{[\mu} G(x-X) M'_{\nu]\lambda}
\end{equation}
where $M'_{\nu\lambda}$ is the magnetic dipole moment of the current distribution,
\begin{equation}
    M'_{\nu\lambda} = \int y_{[\nu} J_{\lambda]}(y) \; d^4y
\end{equation}
and $X$ is some point in the support of $J_\nu$. 

Let us now consider the case of a purely magnetic source.
Since $\del_\mu F''_{\mu\nu}=0$, we may express the dual field strength in terms of a ``dual'' potential $B_\mu$:
\begin{equation}
    \tilde F''_{\mu\nu} = \del_\mu B_\nu - \del_\nu B_\mu
\end{equation}
The Maxwell equations then reduce to
\begin{equation}
    -\del^2 B_\nu = K_\nu
\end{equation}
A literal repetition of the proceeding analysis gives the dual field strength in the form
\begin{equation}
    \tilde F''_{\mu\nu}(x) = 2 \del_\lambda \del_{[\mu} G(x-X) N''_{\nu]\lambda}
\end{equation}
where
\begin{equation}
    N''_{\nu\lambda} = \int y_{[\nu} K_{\lambda]}(y) \; d^4y
\end{equation}
is the ``dual'' magnetic dipole moment of $K_\nu$.
The field strength is thus given by
\begin{equation}
    F''_{\mu\nu}(x) = \epsilon_{\mu\nu\rho\sigma} \del_\lambda \del_\rho G(x-X) N''_{\sigma\lambda}
\end{equation}
Remarkably, this can also be written, after a straightforward computation, in the form
\begin{equation}
    F''_{\mu\nu}(x) = - 2 \del_\lambda \del_{[\mu} G(x-X) \tilde N''_{\nu]\lambda} 
\end{equation}
which is the same formula as in the electric source case, if $M_{\nu\lambda}'$ is substituted for $-\tilde N''_{\nu\lambda}$.

\begin{figure}[t]
\begin{center}
\vspace{-1.2cm}
 \includegraphics[angle=0, width=0.4\textwidth]{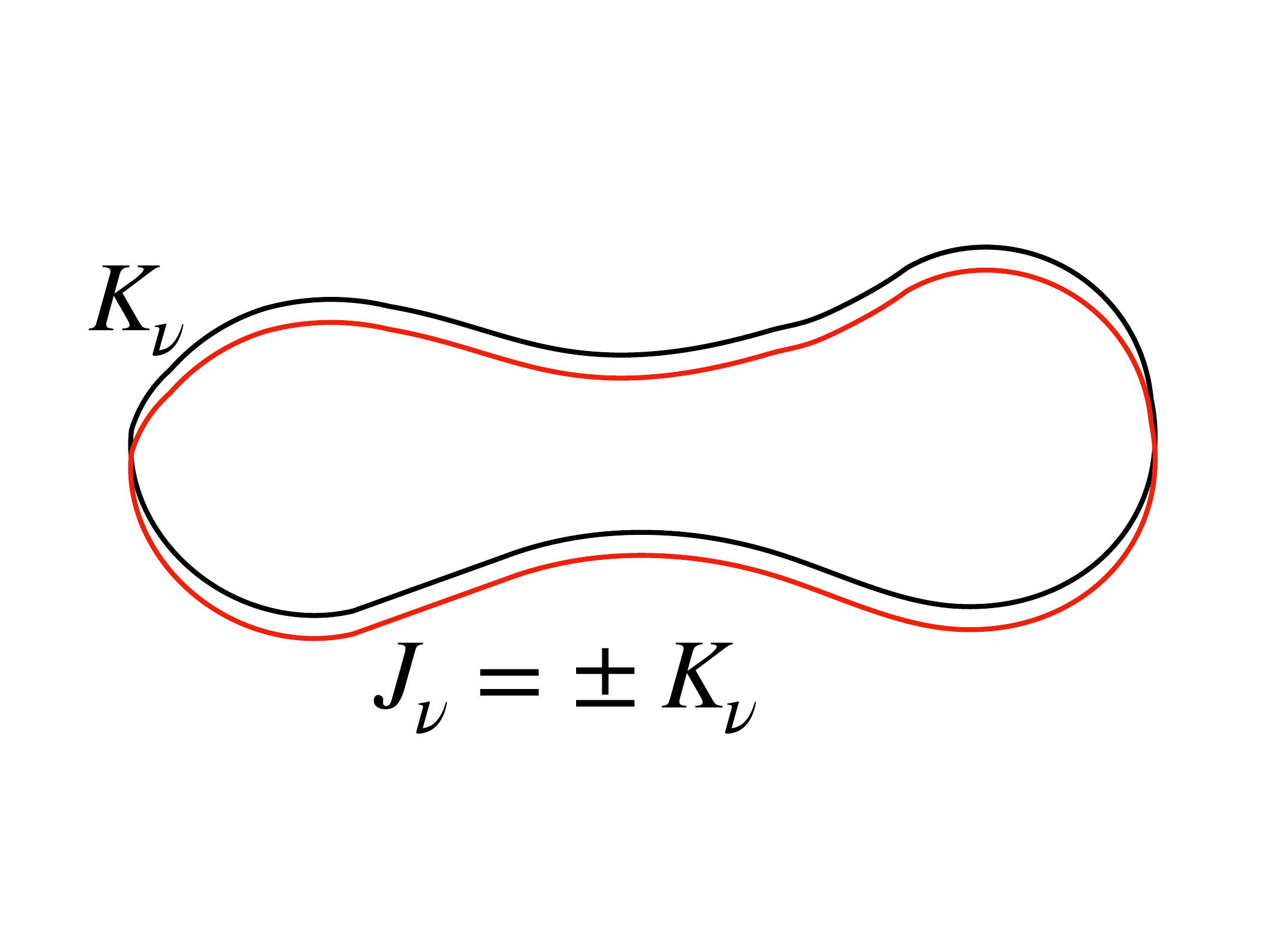}
\vspace{-1cm}
\caption{Euclidean Maxwell equation with a pair of magnetic current  and "electric" current loops  $(K_\nu, J_\nu= \pm K_\nu)$  generates abelian self-dual (or antiselfdual, depending on the sign)  monopole gauge field.  The current lines are on top of each other and are only split for convenience. The self-dual monopole loop is not a dyon loop. See \S.\ref{why}. 4d instanton field at large-distance can be generated by a superposition of  such double-current loops.  }
\label{SD-loops}
\end{center}
\end{figure}

We can now assemble our results.
For a self-dual or anti-self-dual source, we have
\begin{equation}
    N''_{\nu\lambda} = \pm M'_{\nu\lambda}
\end{equation}
and the leading far-field form of the field strength is
\begin{align}
    F_{\mu\nu}(x) &= 2 \del_\alpha \del_{[\mu} G(x-X) M_{\nu]\alpha} \cr
    & = \frac{2}{ \pi^2 |x|^4}( M_{\mu \nu} -2  (M_{\mu \alpha} \hat x_{\alpha}) \hat x_{\nu} + 2   (M_{\nu \alpha} \hat x_{\alpha}) \hat x_{\mu} ) 
    \label{Mon-field}
\end{align}
where we have put
\begin{equation}
    M_{\nu\alpha} = M'_{\nu\alpha} \mp \tilde M'_{\nu\alpha}
\end{equation}
for the net magnetic dipole moment.
Note that a self-dual field arises from an anti-self-dual magnetic dipole moment, and an anti-self-dual field arises from a self-dual magnetic dipole moment.  For example, for self-dual magnetic dipole moment,  we can express  the field strength as a linear combination of anti-selfdual  't Hooft symbols (just like non-abelian instantons): 
\begin{align}
[F_{\mu \nu}] 
= F_{12} \bar \eta^3 + F_{31} \bar \eta^2 + F_{23} \bar \eta^1  
\end{align}
Given  \eqref{Mon-field}, we can easily derive the interactions between small 
monopole loops at large separation.  Let $M^1_{\mu \nu}$  be self-dual and $M^2_{\mu \nu}$ be 
anti-self dual. Then, 
\begin{align}
  & V_{\rm SD-SD} =  V_{\rm ASD-ASD}=0 \cr
   & V_{\rm SD-ASD} =  \frac{2}{ \pi^2 |x|^4} ( -4  (M_{\mu \alpha}^1 \hat x_{\alpha}) (M_{\mu \beta}^2 \hat x_{\beta}))
\end{align}
where the first line follows self-duality of the configurations. Needless to say, these form of interactions is exactly identical between instanton-instanton and instanton-antiinstanton. 

The analogy with the large-distance behavior of $SU(N)$ instantons could hardly be missed.
But we can make the analogy even sharper if we introduce $N^2-1$ copies of Maxwell fields $F^a_{\mu\nu}$, which we are thinking of as the Abelianized components of an $SU(N)$ Yang--Mills field. 
As above, we consider the system with self-dual sources:
\begin{equation}
    -\del_\mu F^a_{\mu\nu} = J^a_\nu,\quad -\del_\mu \tilde F^a_{\mu\nu} = J^a_\nu
\end{equation}
In order to reproduce the large-distance form of the instanton, 
we simply need to choose the currents $J^a_\nu$ so that
\begin{equation}
    M^a_{\nu\alpha} = -8\pi^2 \tr w \bar\sigma_{\nu\alpha} w^\dag T^a
\end{equation}
It is clear that this can always be done.

\subsection{Instantons vs. self-dual monopole loops}
What is this telling us?
At least as far as the large-distance behavior is concerned, the field of an instanton might as well be the field of a collection of   self-dual magnetically charged matter.  
The leading  large distance limit of the instanton field  $\propto (1/|x|^4)$
is reproduced merely in terms of anti-self-dual monopole dipole moments, namely 
\begin{align}
&  A_{\mu}^a =    \frac{M_{\mu \alpha}^a  x_{\alpha}} {|x|^4}, \qquad
F_{\mu \nu}^a = \frac{1}{|x|^4}( M_{\mu \nu}^a -2  (M_{\mu \alpha}^a \hat x_{\alpha}) \hat x_{\nu} + 2   (M_{\nu \alpha}^a \hat x_{\alpha}) \hat x_{\mu} )
\end{align}
where  $(N^2 -1)$  $M^a_{\nu\alpha}$  are   dictated by  $4N-4$ real ADHM moduli data, $w_{u\alpha}$. 
This data leads to the correct large-distance interactions between 4d instantons:
\begin{align}
  & V_{\rm II} =  V_{\rm AA}=0 \cr
   & V_{\rm IA} =  \frac{2}{ \pi^2 |x|^4} ( -4  (M_{\mu \alpha}^{a1} \hat x_{\alpha}) (M_{\mu \beta}^{a2} \hat x_{\beta}))
\end{align}
i.e. instanton-instanton interactions vanishes, and instanton-anti-instanton is the interactions between non-abelian magnetic dipoles. 

Note that we are not using the maximal abelian gauge trick of 't Hooft, which is very widely used in lattice gauge theory. We are also not saying that there are some form of monopole solutions in pure Yang-Mills theory. What we are saying is that, the large distance non-abelian field of an instanton and mutual interactions between instantons can be reproduced by a collection of $N^2-1$ magnetic dipole moments, which are dictated in terms of $4N-4$ moduli of an instanton. 


Nevertheless, if we could obtain a precise quantum mechanical interpretation of the instanton sum as a sum over monopole worldlines, valid at also down to short distances, we would go a long way towards understanding confinement.   However, this cannot be a naive relation. In particular, we have seen that the weight factor for monopole loops of some characteristic size $\rho$ grows exponentially (in lattice formulation), while the weight factor for instantons grows as power law. 
At this stage, we believe that ADHM data  will play a role. 
We can hypothesize that the instanton dipole moments $w\bar\sigma_{\mu\nu}w^\dag$ are generated by some appropriately chosen current loops. 
In this way,
we may still be able to contruct instanton in terms of self-dual magnetic monopoles, but with a highly  orchestrated distribution of self-dual current loops. 
If it really is the case that instantons are certain combination of self-dual monopole worldline loops, then the dominance of arbitrarily large instantons in the Euclidean vacuum can be thought of as the condensation of magnetic monopoles, just as it is in the Abelian lattice gauge theory we have discussed above. 
And once we know magnetic monopoles are condensing, we are done.

\subsection{Is self-dual monopole loop a dyon loop?}
\label{why}
\eqref{MaxwellEM} is the Euclidean Maxwell equation. As shown above, declaring that 
``electric'' current is equal to magnetic current \eqref{MaxSD}, it leads to a self-dual abelian field strength satisfying   the abelian version of instanton equations. In Euclidean and Minkowski spaces, the self-duality equation are, respectively:
\begin{align} 
{\rm Euclidean}    \qquad  E= B \cr 
{\rm Minkowski} \qquad  i E= B
\label{inst}
\end{align} 
In Euclidean space, for a positive action $(S>0)$  configuration, we have to take both  $(E, B)$ real. 
Analogously, for the instanton in non-abelian gauge theory, $F_{\mu \nu}^a$ are all real.  Indeed, with real sources in 
\eqref{MaxwellEM},  we produce a real instanton field in Euclidean space.  
Since both ``electric'' and magnetic current sources are turned on, one may be tempted to think that this self-dual monopole can be interpreted as  a dyon. This is not so.\footnote{  
Below, we wish to explain this as simply as possible. The structure we discuss below is related to a rigorous version of the semi-classical analysis that goes under the name of resurgence and Picard-Lefschetz theory.  To explain it fully would be more technical than we desire in this paper.  A  detailed exposition of the ideas  will appear in a separate work.}

First of all, let us recall that the energy density,  either in the Euclidean space  or in Minkowski space, vanishes for {\bf all} self-dual fields 
\begin{align}
  T^{44}_{\rm Euc} \sim - &E^2 + B^2 =0  \qquad {\rm because}  \qquad   E=B \cr
   T^{44}_{\rm Min} \sim + &E^2 + B^2 =0  \qquad {\rm because}  \qquad   iE=B. 
\end{align}
In fact, all components of the energy=momentum tensor vanish by self-duality (or anti-selfduality), see \cite{Vandoren:2008xg}.  
For a regular dyon  in Minkowski  with real $(E, B)$ fields,  obviously energy density is positive.  

The definition of the electric and magnetic charge in Minkowski space follows from Gauss' law:
\begin{align}
    \int_{S^2_{\infty}} \vec E . d \vec S  = \int J_0  \in e \Z,  \cr
    \int_{S^2_{\infty}} \vec B . d \vec S  =  \int K_0 \in g \Z, 
    \label{Gauss}
\end{align}
and $(e, g)$ satisfies Dirac quantization condition $eg= 2 \pi n$. Needless to say, both charges are real.    Below, we determine the sources for various self-dual fields. The outcome is surprising. 

Let us start with purely real $(E, B)$ saddle point in Euclidean space. 
Note that in Minkowski space,  $B$ associated with instanton is real,  while $E$ is purely imaginary.  If we were to ask what type of sources would generate such a field in Minkowski space, we would conclude that magnetic charge must be real, but electric charge must be imaginary, both being measured by using Gauss' law.   Furthermore, we would discover that both magnetic charge as well as imaginary electric charge are in units of $g$. This configuration  is the Wick rotation of the self-dual monopole field from Euclidean space to Minkowski space. In other word, self-dual completion of the monopole field requires, apart from the real monopole charge, an imaginary electric charge, that does not appear in \eqref{Gauss}.  See Table.\ref{SD-completion}.

\begin{table}[h]
\begin{center}
\begin{tabular}{ |c|c|c|c|c|c|c|c| } 
 \hline
 Metric & $\vec E$  &  $\vec B$    &  $J^e_{0}$  &  $J^e_{i}$   &$K^m_{0}$  &  $K^m_{i}$  & SD?   \\ 
\hline
 \hline
 \hline
 Euc& $\R$  &  $ \R$    &   $g \Z$    & $g \Z$    & $g \Z$    & $g \Z$    & SD mon.    \\ 
 \hline 
  Min& $i \R$  &  $ \R$    &   $g i  \Z$    & $g \Z$    & ${\color{red} g \Z}$    & $g i  \Z$    & SD mon.   \\
  \hline
 \hline \hline
 Euc& $i \R$  &  $ i \R$    &   $e i  \Z$    & $e i  \Z$    & $e i  \Z$    & $e i \Z$    
 & SD e-charge   \\ 
 \hline 
  Min& $\R$  &  $ i \R$    &   $ {\color{red}e  \Z} $    & $e i \Z$    & ${e i \Z}$    & $e  \Z$    & SD e-charge \\
    \hline
 \hline \hline
 Euc& $\R$  +   $i \R$   &  $ \R$  +  $i \R$    &   $g \Z$   + $e i  \Z$     & $g \Z$   + $e i  \Z$     & $g \Z$   + $e i  \Z$     & $g \Z$  + $e i  \Z$     & SD dyon    \\ 
 \hline 
  Min& $i \R$   +  $  \R$   &  $ \R$  +  $i \R$    &   $g i  \Z$  + ${\color{red}e  \Z}$    & $g \Z$  +  $e i  \Z$    & ${\color{red} g \Z}$  +  $e i  \Z$    & $g i  \Z$   +  $e  \Z$  
  & SD dyon   \\
  \hline
 \hline \hline
  Euc& $ i \R$  &  $ \R$    &   $e i \Z$    & $e \Z$    & $g \Z$    & $i g \Z$    & not SD, dyon   \\ 
 \hline 
  Min& $ \R$  &  $ \R$    &   $e  \Z$    & $e \Z$    & $ g \Z$    & $g   \Z$    & not SD,dyon  \\
  \hline
 \hline \hline
 \end{tabular}
\end{center}
\vspace{0.5cm}
\caption{Wick rotations of self-dual fields and the structure of currents sourcing them in both Euclidean and Minkowski space.   $\vec B,  J^e_{i}, K^m_{0}$ remains  the same, while  
$\vec E,  J^e_{0}, K^m_{i}$ are multiplied with either  $\pm i$ under Wick rotation.
First two input  lines are for self-dual monopoles, third and fourth are for self-dual electric-charged particles, and fifth and sixth are for self-dual dyon. Last two lines are for ordinary (non-selfdual) dyon configuration. Self-dual configurations have $T^{44}=0$ while an ordinary dyon has $T^{44}>0$.}
\label{SD-completion}
\end{table}

The self-duality equation in Euclidean space also possesses purely imaginary $(E, B)$ solutions. However,  these would be 
negative action configurations $(S<0)$, similar to ghost-instantons in quantum mechanics \cite{Basar:2013eka, Dunne:2015eaa}.
Their Stokes multipliers should be strictly zero 
for a meaningful semi-classical expansion.  Yet, the Wick rotation of this field to Minkowski space leads to a real $E$ field, and imaginary 
$B$ field. If we ask what type of sources would induce such fields in Minkowski space, we learn that real electric charge and imaginary magnetic charge configurations do.  Thus, the self-dual completion of the electrically charged particle is achieved with an addendum of imaginary  magnetic charge.

The self-duality equation in Euclidean space also possess  solutions solutions for which   $(E, B)$ are complex, with non-zero real and imaginary parts.  This can be viewed as a superposition of the above two constructions. Whether action is positive or negative depends on the charges.  In Minkowski space, $E= -i B$. If we ask what type of charges in Minkowski space sources such fields,  the real parts certainly require the real electric and  magnetic charges, but we must also have imaginary electric and imaginary addendums to achieve self-duality. This is the self-dual version of the dyon.

A regular dyon in Minkowski space is only sourced by real $(e, g)$ pair, with no imaginary charges. As such, it is not self-dual. 

 At a formal level, the saddles in the problem we have are solutions to (complexified) self-duality equation, where real $F_{\mu \nu}$ is promoted to a $\mathbb C$-valued  field strength ${ \cal F}_{\mu \nu}$,  
 \begin{align}
{ \cal F}_{\mu \nu} = * { \cal F}_{\mu \nu}
\label{inscom}
 \end{align}
 with both real and imaginary solutions.\footnote{Although this may seem exotic, modern treatment of semiclassics based on Picard-Lefshetz theory tells us that we have to start with complexified versions of equations of motions \cite{Dunne:2015eaa}.  This is true for  steepest descent method
applied to finite dimensional integrals as well \cite{Witten:2010cx}. In fact, 
$\mathbb C$-valued  field strength ${ \cal F}_{\mu \nu}$ also appears in analytic continuation of Chern-Simons theory \cite{Witten:2010cx}.
Here, we rediscovered the necessity of complexification with reverse rationale. We have seen that a self-dual dyon generates necessarily a complex field  configuration either in Euclidean or Minkowski space, while an ordinary dyon produce real $(E,B)$ field in Minkowski space.} 
 Equivalently, when we think of Maxwell's equation to build the saddles, we have to view both sources and fields as complexified as well: 
 \begin{equation}
    -\del_\mu {\cal F}_{\mu\nu} = {\cal J}_\nu,\quad -\del_\mu \tilde {\cal F}_{\mu\nu} = {\cal K}_\nu 
    \label{MaxwellEMcom}
\end{equation}
For clarity, let us consider Minkowski space. Let $(\cal E, \cal B)$ denote complexification of $(E,B)$. Thus,  we can  express the form of the charges leading to self-dual fields  in Minkowski space as 
 \begin{align}
 \int_{S^2_{\infty}} \vec {\cal E} . d \vec S   &= \int  {\cal J}_0  \in   e \Z  +  g i  \Z, \\
 \int_{S^2_{\infty}} \vec {\cal B} . d \vec S   &= \int  {\cal K}_0    \in g \Z  +  e i  \Z  
    \label{MaxwellEMcom}
\end{align}
which should be viewed as generalization of Gauss' law \eqref{Gauss} to self-dual fields. 
The Dirac quantization $eg= 2 \pi \Z$ is associated with real parts of the charges. 
The structure emanating from instanton  are self-dual monopoles, with $e=0$ in \eqref{MaxwellEMcom}. In other words,  ${\rm Re}{\cal J}_0 =0$, and ${\rm Im}{\cal J}_0 \propto g$ must be viewed as part of  self-dual completion of monopole field. 

To summarize, let us provide a list of charges sourcing  ordinary real  electric and magnetic fields, and charges sourcing  self-dual field configurations:  
\begin{align}
   \left( \int  {\cal J}_0,  \int  {\cal K}_0 \right) = \left\{ \begin{array}{ll}
      (0, gn)  &  {\rm Monopole}   \\ 
   (em, 0)     &  {\rm e-charged\;  particle} \\ 
    (em, gn)  &  {\rm Dyon}  \\ \\
    (ign, gn)  &  {\rm SD \; Monopole}  \\
   (em, iem)     &  {\rm SD\; e-charged\;  particle} \\
    (em + ign, gn +iem)  &  {\rm SD \; Dyon}  
   \end{array} \right.
\end{align}
where $n,m \in \Z$. Clearly, there is some room to confuse a dyon with a self-dual monopole. But in Minkowski space, where we define charges via Gauss' law, the e-charge for dyon is real and in units of $e$, while the e-charge of self-dual monopole is imaginary and in units of $g$.

\section{The transmutation of instanton operator: $\R^4$  vs.  small $\R^3 \times S^1$}
\label{sec:trans}

Below, we would like to show that the instanton operator that we write down on $\R^4$ transmutes to the instanton operator that we would 
obtain on  small $\R^3 \times S^1$ from monopole operators. To this end, let us first show what to expect on  $\R^3 \times S^1$, starting with the monopole constituents of an instanton. Then, starting with the instanton operator on $\R^4$, we obtain the same operator on $\R^3 \times S^1$ by invoking abelianization due to gauge holonomy and abelian duality.   In translating the instanton moduli on  $\R^3 \times S^1$ from the ADHM parametrization to the monopole moduli, Hopf map plays a crucial role.

\subsection{Abelian duality on the instanton operator}

Let us start with pure $U(1)$ gauge theory in 4d. 
We write it in a formulation that is intermediate between the original and the dual ones as in \cite{Witten:1995gf}.
This intermediate formulation involves two 1-form gauge fields $A,B$ and a 2-form gauge field $H$, and the action is given by
\begin{equation}
    S = \frac14 \int (F_{\mu\nu} + H_{\mu\nu})^2 - \frac{i}{2} \int \tilde H_{\mu\nu} G_{\mu\nu}
\end{equation}
where $F,G$ are the field strengths of $A,B$, respectively.%
\footnote{
In this section, since our concerns are mainly kinematic, we scale the fields to absorb the gauge coupling and simplify various factors. 
}
Indeed, it is easy to see that integrating out $B$ and $H$ results in the original Maxwell theory, whereas integrating out $A$ and $H$ results in Maxwell theory with $B$ as the fundamental field. 
Let us show that for any antisymmetric 2-index tensor $M_{\mu\nu}$, not necessarily self-dual or anti-self-dual, the operator $\hat{\mathcal I}$ defined in this intermediate formulation by 
\begin{multline}
    \langle \widehat{\mathcal I}(X) \rangle 
    = \int (dA) (dB) (dH) \exp \biggl\lbrack -\frac14 \int (F_{\mu\nu}+H_{\mu\nu})^2 - \frac{i}{2} \int \tilde H_{\mu\nu} G_{\mu\nu} \biggr\rbrack \\
    \times \exp\{ \tfrac14 M_{\mu\nu} \lb F_{\mu\nu}(X) + i G_{\mu\nu}(X)\rb \}
\end{multline}
is completely equivalent to the instanton operator in the original formulation of $U(1)$ gauge theory.
This operator can also be expressed using the magnetic dipole moment density $m_{\mu\nu}(x) \equiv M_{\mu\nu} \delta^4(x-X)$ as
\begin{equation}
    \langle \widehat{\mathcal I}(X) \rangle 
    = \int (dA)(dB)(dH) \; \exp \biggl\lbrack - \frac14 \int (F_{\mu\nu} + H_{\mu\nu})^2 - \frac{i}{2} \int H_{\mu\nu} \tilde G_{\mu\nu} + \frac14 \int m_{\mu\nu} (F_{\mu\nu} + i G_{\mu\nu}) \biggr\rbrack
\end{equation}
On integrating out $B$, the linear terms in $B$ inside the exponential are converted into delta-functional factor under the path integral sign:
\begin{equation}
    \langle \widehat{\mathcal I}(X) \rangle 
    = \int (dA)(dH) \; \exp \biggl\lbrack - \frac14 \int (F_{\mu\nu} + H_{\mu\nu})^2 + \frac14 \int m_{\mu\nu} F_{\mu\nu} \biggr\rbrack \; \delta \lb \del_\mu \tilde H_{\mu\nu} - \tfrac12 \del_\mu m_{\mu\nu} \rb 
\end{equation}
Ignoring topological subtleties, the delta functional localizes the $H_{\mu\nu}$ integration to $H_{\mu\nu}=\tfrac12 \tilde m_{\mu\nu}$.
Thus,
\begin{equation}
    \begin{split}
    \langle \widehat{\mathcal I}(X) \rangle 
    &= \int (dA) \; \exp \biggl\lbrack - \frac14 \int (F_{\mu\nu} + \tfrac12 \tilde m_{\mu\nu})^2 + \frac14 \int m_{\mu\nu} F_{\mu\nu} \biggr\rbrack \\
    &= \int (dA) \; \exp \biggl\lbrack - \frac14 \int (F_{\mu\nu})^2 - \frac{1}{16} \int (\tilde m_{\mu\nu})^2 - \frac14 \int \tilde m_{\mu\nu} F_{\mu\nu} + \frac14 \int m_{\mu\nu} F_{\mu\nu} \biggr\rbrack \\
    &\propto \int (dA) \; \exp \biggl\lbrack -\frac14 \int (F_{\mu\nu})^2 + \frac14 \int (m_{\mu\nu} - \tilde m_{\mu\nu})F_{\mu\nu} \biggr\rbrack \\
    &= \int (dA) \; \exp \biggl\lbrack -\frac14 \int (F_{\mu\nu})^2 \biggr\rbrack \exp \lb \tfrac12 M_{\mu\nu}^- F_{\mu\nu}(X) \rb
    \end{split}
\end{equation}
where $M_{\mu\nu}^-$ is the anti-self-dual part of $M_{\mu\nu}$ and we have dropped the (divergent) contact term $\int (\tilde m_{\mu\nu})^2$ which can be absorbed into the normalization of $\widehat{\mathcal I}$.
Thus, we have found
\begin{equation}
    \langle \widehat{\mathcal I} \rangle_{\text{intermediate}} = \langle \mathcal I \rangle_{\text{original}}
\end{equation}
as claimed. 

We can easily generalize this discussion to case of $N^2-1$ copies of Maxwell fields.
In doing so, we may express the Yang--Mills instanton operator in the following form.
Let $M_{\mu\nu}^a$ ($a=1,\ldots,N^2-1$) be a collection of constant antisymmetric tensors such that
\begin{equation}
    M_{\mu\nu}^{a-} = -8\pi^2 \tr \lb w \bar\sigma_{\mu\nu} w^\dag  T^a \rb
\end{equation}
Then we find that the instanton operator 
\begin{equation}
    \mathcal I = \exp \lb \tfrac12 M_{\mu\nu}^{a-} F^a_{\mu\nu} \rb
\end{equation}
where $M_{\mu\nu}^{a-}$ is the anti-selfdual part of $M_{\mu\nu}^{a}$, can be expressed as
\begin{multline}
    \langle \widehat{\mathcal I}(X) \rangle 
    = \int (dA)(dB)(dH) \exp\biggl\lbrack -\frac14 \int (F_{\mu\nu}^a +H_{\mu\nu}^a )^2 -\frac{i}{2} \int \tilde H_{\mu\nu}^a G_{\mu\nu}^a \biggr\rbrack \\
    \times \exp\lb \tfrac14 M_{\mu\nu}^a (F_{\mu\nu}^a + i G_{\mu\nu}^a)  (X) \rb
\end{multline}

\subsection{3d dimensional reduction of 4d Abelian duality}

Let us consider the dimensional reduction to 3d of the self-dual formulation of 4d pure Maxwell theory.
We begin by explicitly separating out the components in the Lagrangian as follows and assuming all fields are independent of the fourth coordinate $x_4$: 
\begin{equation}
    L = \tfrac12 (\del_i A_4 + H_{i4})^2 + \tfrac14 (F_{ij} + H_{ij})^2 + \tfrac12 i \epsilon_{ijk} H_{ij} \del_k B_4 + i \epsilon_{ijk} H_{i4} G_{jk}
\end{equation}
Integrating out $B_k$ imposes the constraint $\epsilon_{ijk}\del_j H_{i4} =0$.
Ignoring topological subtleties, this implies $H_{i4}$ is a gradient of a scalar $\alpha$, $H_{i4} = \del_i \alpha$.
Making this substitution above, we get
\begin{equation}
    L= \tfrac12 (\del_i A_4+\del_i\alpha)^2 + \tfrac14 (F_{ij} + H_{ij})^2 + \tfrac12 i \epsilon_{ijk} H_{ij} \del_k B_4 
\end{equation}
Next, we make the field redefinitions $A_4 \to - \alpha$ and $H_{ij} \to H_{ij}-F_{ij}$ and then integrate out $H_{ij}$.
This result is the Lagrangian
\begin{equation}
    L = \tfrac12 (\del_i A_4)^2 + \tfrac12 (\del_k B_4)^2 
\end{equation}
We typically put $A_4 = \phi$, $B_4=\sigma$, so this is rewritten as
\begin{equation}
    L = \tfrac12 (\nabla \phi)^2 + \tfrac12 (\nabla \sigma)^2
\end{equation}
Clearly, this derivation generalizes to the case of several copies of Maxwell theories.
We then interpret $\phi$ and $\sigma$ in the final answer as multicomponent scalar fields, $\phi=(\phi_1,\ldots,\phi_n)$ and $\sigma=(\sigma_1,\ldots,\sigma_n)$.

\subsection{Instanton operator on $\R^3 \times S^1$ from monopole operators: $SU(2)$ case } 
 
In $SU(2)$ gauge theory on small $\R^3 \times S^1$ that undergoes abelianization due to a center-symmetric gauge holonomy, there are two types of monopoles, BPS and KK. 
Let us denote their respective operators by ${\cal M}_1 (x_1)$ and  ${\cal M}_2 (x_2)$, where $x_1, x_2 \in \R^3$:
\begin{align}
{\cal M}_1(x_1) = e^{- \phi(x_1) + i \sigma(x_1) }  =   e^{- z(x_1) }   \qquad {\cal M}_2(x_2) = e^{+ \phi(x_2) - i \sigma(x_2) }    = e^{+z(x_2) }  
\end{align}  
Here, $\phi$ is the fourth component of the  gauge potential, $A_4$ in the Cartan subalgebra direction, and $\sigma$ is the scalar field dual (in the three-dimensional sense) to the 3-vector part of the potential $A_i$.  
Note that the representation of the magnetic Coulomb charges in the dual representations are of the form  $e^{i q \sigma(x) }$, so 
${\cal M}_1$ represents a charge $+1$  monopole, ${\cal M}_2$ a charge $-1$ monopole.  
The coefficient $q_{\rm dil}$ in $e^{- q_{\rm dil} \phi(x) }$ can be viewed as a dilaton charge (and crucially not the electric charge), associated with the spontaneous breaking of dilatation symmetry by the adjoint Higgs vev \cite{Harvey:1996ur}. 
The standard electric charge would couple to the gauge holonomy field as  $e^{ i q_{\rm e} \phi(x) }$.%
\footnote{
This is related to the reason we put quotes around ``electric'' in reference to the source of $\del_\mu F_{\mu\nu}$ in the previous section. 
}
If one wishes to interpret the charge coupling to the gauge holonomy part of the monopole operator in this language, (which is useful to formulate a statistical mechanics based on proliferating monopoles), it must be viewed as an imaginary electric charge, at best. 
In this sense, the Coulomb  charges appearing in the monopole operator can be viewed as a pair of imaginary and real charges :
\begin{equation}
    \begin{aligned}
    &\text{monopoles}\colon  &&q_1= (+i, +1),  &&q_2 = (-i, -1)   \\ 
    &\text{antimonopoles}\colon  &&q_1= (+i, -1), &&q_2 = (-i, +1) 
    \end{aligned}
\end{equation}
The self-duality of the monopoles requires the Coulomb interaction between self-dual objects to vanish, while the interaction between a selfdual and antiselfdual monopole falls off as $1/r$. 
The classical interactions between monopoles are 
\begin{align}
 \big\langle  {\cal M}_1  (x) {\cal M}_1 (y)   \rangle &= e^{- (   \frac{i^2}{|x-y|}    + \frac{1^2}{|x-y|} )} = e^0,  \cr 
 \big\langle  {\cal M}_1  (x) {\cal M}_2 (y)  \big\rangle  &=  e^{- (   \frac{i (-i)}{|x-y|}    + \frac{1 (-1)}{|x-y|} )} = e^0 \cr 
  \big\langle  {\cal M}_1  (x) \bar {\cal M}_1 (y)   \rangle &= e^{- (   \frac{i^2}{|x-y|}    + \frac{1 (-1)}{|x-y|} )} = e^{ - (- \frac{2}{|x-y|})}  \cr
    \big\langle  {\cal M}_1  (x) \bar {\cal M}_2 (y)   \rangle &= e^{- (   \frac{i (-i)}{|x-y|}    + \frac{(1)^2}{|x-y|} )} = e^{ - ( \frac{2}{|x-y|})}  
 \end{align} 
Recall that on small $\R^3 \times S^1$, an instanton is a composite of ${\cal M}_1(x_1)$  and ${\cal M}_2(x_2)$. 
Let us call the center position of monopoles $x_{\rm c}$, and the relative coordinates  with respect to center $\xi_1,\xi_2$. Hence, 
\begin{equation}
    x_{\rm c}= \frac{1}{2} (x_1 + x_2),  \qquad \xi_1 = x_1 -x_{ \rm c},  \;\;  \xi_2 = x_2 -x_{ \rm c}
\end{equation} 
Let us consider an instanton of fixed size, centered at  $x_{\rm c}$. 
We can expand the field $z(x_i)$ around the center position $x_{ \rm c}$ as  $z( x_{\rm c} +\xi_i) = z( x_{ \rm c}) +  (\xi_i\cdot \nabla) z(x_{\rm c}) + \ldots $
\begin{equation}
    \begin{split}
    {\cal I}  &=  {\cal M}_1 (x_1)   {\cal M}_2  (x_2)  = e^{- z(x_1) }  e^{z(x_2) }    \\     &\approx e^{- (\xi_1 - \xi_2) \cdot \nabla z (x_{\rm c}) }    =  e^{- d\cdot \nabla \phi  (x_{\rm c}) + i  d \cdot \nabla \sigma  (x_{\rm c})  }
    \label{Instsu2}
    \end{split}
\end{equation}
We observe that we have two dipoles, one imaginary (associated with $\phi$) and one real (associated with $\sigma$), following from our charge assignments for monopoles.   
We view the instanton on small $ \R^3 \times S^1$ as a pair of dipoles, 
\begin{equation}
    \begin{aligned}
    &\text{instanton}\colon && (i \vec d, +\vec d)  \\
    &\text{anti-instanton}\colon && (i \vec d',   - \vec d') 
\label{Instsu2-int}
    \end{aligned}
\end{equation}
where $\vec d \equiv \vec \xi_1-\vec\xi_2$.
This charge assignment guarantees that  the interaction between two instantons (or two anti-instantons) is zero because $(\pm i)^2 +1=0$, while the interaction between an instanton and anti-instanton is of the dipole--dipole type in 3d, decaying as ${1}/{r^3}$. 
Note that these formulas are valid for fixed size instantons at asymptotically large separation.  
Since the interaction between constituents ${\cal M}_1, {\cal  M}_2$ of an instanton vanishes, it is more appropriate to view the instanton as   {\it dipoles at threshold}, rather than {\it permanent dipoles}, because at the classical level, there is no binding between constituents. 
In other words, the modulus of the dipole vector 
$ |\vec d| $ is unbounded from above. It is associated with the size parameter $\rho$ of the 4d instanton, the relation being $ |\vec d| L = \rho^2$ as proven below, where $L \lesssim \Lambda^{-1}$ is the length of $S^1$ in the $\R^3 \times S^1$ compactification.

\begin{figure}[t]
\begin{center}
 \includegraphics[angle=0, width=0.8\textwidth]{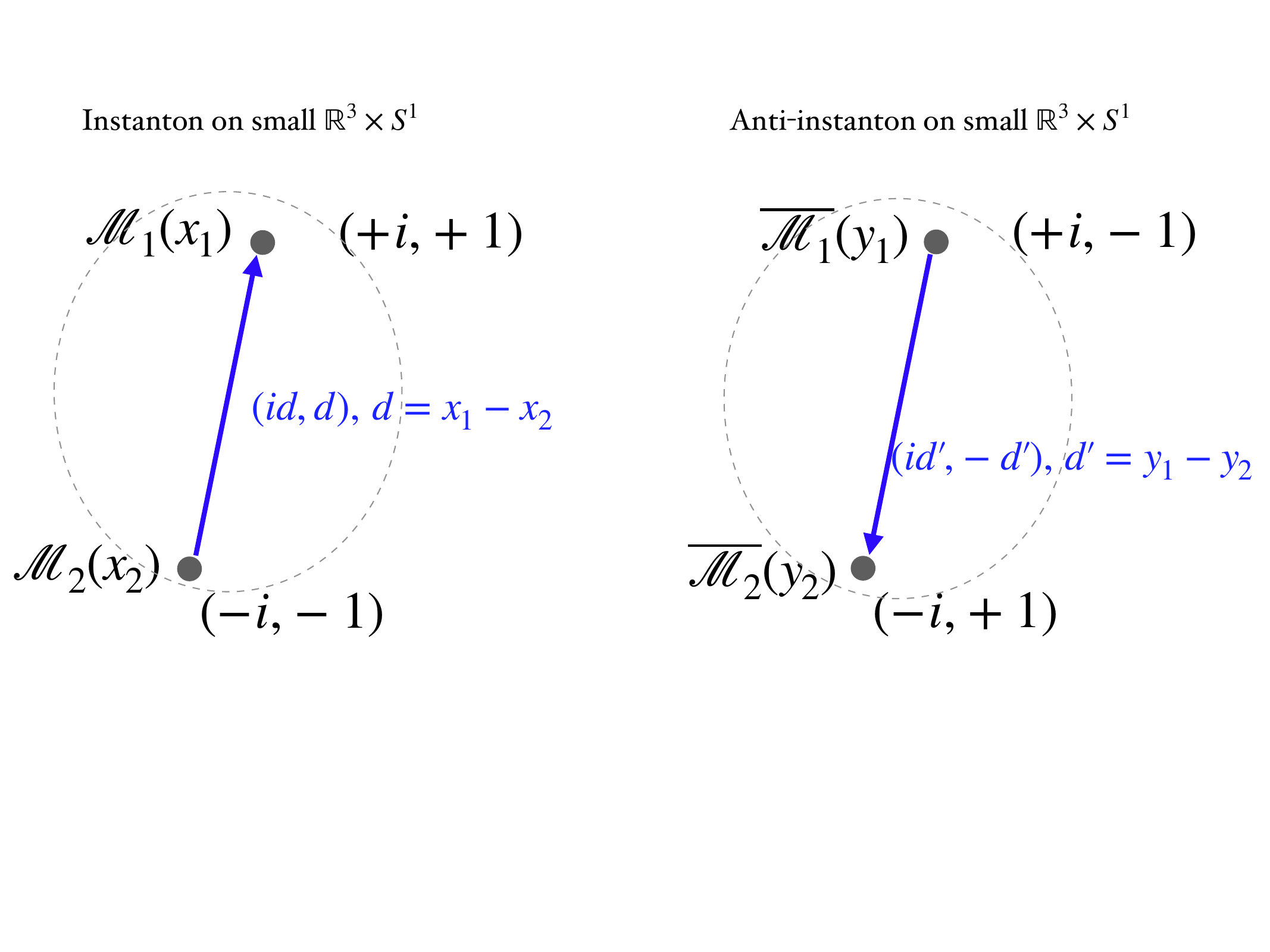}
\vspace{-2.8cm}
\caption{ The instanton operator on small $\R^3 \times S^1 $ \eqref{Instsu2}, pictorially. At the classical level, an $SU(2)$ instanton  
 is a combination of a real dipole and imaginary dipole. This structure 
 is respectful to the  classical moduli space structure, and classical interactions  of both monopoles and instantons.
The 4d instanton operator, upon dimensional reduction (using center-symmetric gauge holonomy and abelian duality), matches to the one obtained from monopole-instanton constituents 
 on $\R^3 x S^1 $ as shown below.  
      }
\label{double-dipole}
\end{center}
\end{figure}

\subsection{\texorpdfstring{From instanton operator on $\R^4$ to the one in $\R^3 \times S^1$: $SU(2)$ case}{From instanton operator on R4 to the one on R3xS1 at long distances}}
Let us re-express  the instanton operator in 4d in terms of chromo-electric and chromo-magnetic field strength components:
\begin{equation}
    \mathcal I 
 =  \exp\{-  \tr \lb w  \vec\tau  w^\dag \cdot (\vec E + \vec B) \rb  \} 
\end{equation}
As before, $w_{u \dot \alpha}, \dot \alpha=1,2, \; u=1, \ldots N$ provides a parametrization of the centered moduli space of unit instantons in $SU(N)$ gauge theory. 
It is described by $2N$ complex ($4N$ real)  parameters obeying  four real ADHM constraints. 
Hence, $w_{u \dot \alpha}$ provides a parametrization of the $4N-4$ dimensional centered moduli space.  
Here, we will focus on gauge group $SU(2)$. 


For gauge group $SU(2)$, $w$ is simply a two-by-two matrix subject to the ADHM constraint $w^\dag w \propto 1_2$.
This means that up to a scale factor, $w$ is a unitary matrix.
The most general form of such a matrix is given by
\begin{equation}
    w =  e^{i\theta}\begin{pmatrix}
        \a & \b \\
        -\b^* & \a^*
    \end{pmatrix}, \qquad \det w = \rho^2
\end{equation}
with $\rho >0$, $\theta$ real, and $\a,\b$ complex numbers such that $|\a|^2 + |\b|^2 = \rho^2$.
Actually, $\theta$ is a redundant parameter -- $w$ only enters the instanton solution via $w\bar\sigma_{\mu\nu}w^\dag$ -- and we can simply set $\theta=0$. 

With this input, the non-abelian  magnetic moments for the $SU(2)$ gauge theory can be written as:
  \begin{align}
     (w \tau_{1} w^\dag ) &=    \begin{pmatrix}
        \a \b^* +   \b \a^*  &   \a^2 -   \b^2   \\
    \a^{*2} -  \b^{*2}   &     -(\a \b^* +   b \a^*)
     \end{pmatrix}  
  \nonumber \\
     (w \tau_{2} w^\dag )&=    \begin{pmatrix}
        \a \b^* -   \b \a^*  &   \a^2 +   \b^2   \\
   - \a^{*2} -  \b^{*2}   &     -(\a\b^* -   \b \a^*)
     \end{pmatrix}   (-i) \nonumber\\
     (w \tau_{3} w^\dag)&= 
     \begin{pmatrix}
      |\a|^2 -  |\b|^2  &   -2 \a  \b     \\
        -2  
        \a^*  \b^*    &  -(|\a|^2 -  |\b|^2)   
     \end{pmatrix}   
    \label{moments-su2-sim}
\end{align}
Obviously, these matrices are not simultaneously diagonalizable. However, they share  (as they must) the same eigenvalues given by  $\pm (  |\a|^2 +  |\b|^2 )  = \pm \rho^2$ where $\rho$ is the size parameter of an instanton.

On small $\R^3 \times S^1$ adiabatically connected to $\R^4$, the $SU(2)$ gauge group always abelianizes to $U(1)$ due to gauge holonomy in the compact direction, and only gauge field components along the Cartan subalgebra, i.e, the gauge fields in the $T^3$ direction, survive at long-distances.
Therefore, in the instanton operator,  $\exp \{-  \tr \lb w  \tau^a  w^\dag T^3 \rb (E_a^3 + B_a^3)  \}$   only the magnetic dipoles in the  Cartan $T^3$ direction  survive.  
Due to the Abelianization, we can formally set the non-Cartan components of $\vec B$  and $\vec DA_4$ to zero.
The instanton operator then acquires the form
\begin{equation}
    \mathcal I = \exp \{ - L\vec d \cdot (\vec DA_4^3 + \vec B^3) \} 
\end{equation}
where we introduced the 3d dipole vector
\begin{equation}
    \vec d = L^{-1} \tr\lb w \vec\tau w^\dag T^3 \rb 
\end{equation}
and we recognize the combination $\vec DA_4^3 + \vec B^3$
from the BPS equation in 3d. 
Using abelianization $L\vec D A_4^3 \rightarrow \vec \nabla \phi$, and abelian duality 
$L\vec B^3  \rightarrow  -i \vec\nabla \sigma$, we can express the instanton operator at long-distances on $\R^3 \times S^1$ as:
\begin{align}
\mathcal I  & =  \exp \{- \vec d  \cdot (  \vec\nabla \phi   -    i \vec\nabla \sigma ) \}  = \exp \{-   \vec d  \cdot \vec \nabla z  \}, \qquad z \equiv \phi -i \sigma
\label{inst3}
\end{align} 
which is identical to \eqref{Instsu2} that we obtained by using monopole constituents of an instanton on small $\R^3 \times S^1$.
To reiterate \eqref{Instsu2-int}, this operator can be viewed as a pair of dipoles, a real dipole associated with $\sigma$  and  an imaginary dipole associated with $\phi$. 

Normalizing the Cartan generator as $T^3 = \operatorname{diag}(1,-1)$, the explicit correspondence between the moduli  parameters entering ADHM and the physical dipole on  $\R^3 \times S^1$ regime is given by
\begin{align}
Ld^1 =  &   \a \b^* +   \b \a^*  = 2 {\rm Re } ( \a \b^*)   \cr
Ld^2= &       -i (\a \b^* -   \b \a^*) = 2 {\rm Im } ( \a \b^*) \cr 
 Ld^3= &         \a \a^* -   \b \b^*,
     \label{moments}
\end{align}
The modulus of  the dipole moment is 
\begin{align}
 |\vec d| =  
   L^{-1} \rho^2 = L^{-1}  ( |\a|^2 +   |\b|^2), \qquad 
     \label{dipolelengths}
\end{align}
For fixed $\rho>0$, we recognize the correspondence taking the ADHM parameters $(\a,\b)$ to the dipole moment $\vec d$
\begin{align}
S^3 \rightarrow  S^2 \qquad (\a,\b) \mapsto \vec d
     \label{moments}
\end{align}
as the Hopf map,
where the $S^3$ has radius $\rho$ and the $S^2$ as radius $\rho^2/L$. 

To summarize the matching: the center moduli of the instanton  $X \in \R^4$   in 4d reduces to $x_{\rm c} \in \R^3 $ (the center position of $({\cal M}_1, {\cal M}_2)$ 
pair and an internal angular moduli  (which may be interpreted as the center position of the instanton in the $S^1$ direction).    
The size    of the 4d instanton reduce to the (square root of the) modulus of the dipole vector  built from    $({\cal M}_1, {\cal M}_2)$ 
pair. Finally, angular $S^3$ moduli  reduce to the orientation of the $\vec d= \xi_1 - \xi_2 \in  S^2$ and the relative internal angular moduli. In other words, the preimage of a  fixed point $\vec d  \in S^2$   is a closed loop in $S^3$. This  closed loop is parametrized by  the  relative internal angular moduli  of the  $({\cal M}_1, {\cal M}_2)$  pair.

\section{Conclusions and Outlook}

For a long time now,  the widely accepted  perspective in Yang-Mills theory and QCD  on $\mathbb R^4$ is that  instantons are not responsible for confinement. 
Rather, either condensation of monopoles or proliferation of center-vortices are viewed as plausible scenarios, the latter especially by the lattice gauge theory community.    
However, the idea of adiabatic continuity challenged this perspective over the last several years. 
If we construct the compactifications on $\mathbb  R^3 \times S^1$ or on $\mathbb  R^2 \times T^2$ with 't Hooft flux, both having the merit that the phase structure in these calculable regimes are continuously connected to the strongly coupled theory on $\mathbb  R^4 $, 
fractional instantons  with topological charge $Q= \frac{1}{N}$  play prominent roles in generating confinement, mass gap and fractional theta dependence \cite{Unsal:2008ch, Unsal:2007jx, Tanizaki:2022ngt, Poppitz:2021cxe}.  
 These  fractional instantons  are monopole-instantons in the former set-up  \cite{Kraan:1998sn, Kraan:1998pm, Lee:1997vp, Lee:1998bb, Lee:1998vu} and  center-vortices in the latter \cite{Tanizaki:2022ngt},  and furthermore, the two types of structures are continuously connected \cite{Hayashi:2024new, Guvendik:2024umd} to each other. 
 This tells us that the internal structure of instantons, with the appropriate conditions, {\bf is}   responsible for confinement in these semi-classical regimes adiabatically connected to the theory on  $\mathbb R^4$.  
 Even this remarkable fact necessitates a more dedicated study of instantons on $\mathbb  R^4 $.  
 Perhaps there is something about  them much deeper than  meets the eye. 

Another crucial progress comes from the general study of phases of theories with $\mathbb Z_N$ 1-form symmetry, starting with a topological BF theory and incorporating the deformation class idea \cite{Nguyen:2024ikq}.  
This study shows that  proliferation of center vortices is necessary but insufficient for confinement, while  the  proliferation of magnetic monopoles is  both necessary and sufficient to induce confinement.
However, it should also be kept in mind that monopoles cannot proliferate unless center vortices do.  
In calculable models, proliferation of the vortices leads to a transition from topological BF phase (also called Higgs phase)  to Coulomb phase, and proliferation of monopoles leads to a transition from Coulomb phase to confining phase.

To us,  it seems very plausible that the internal structure of the 4d instanton  
carries structures that are much more sophisticated than ever thought.  
With this perspective, we reinvestigated some aspects of the $SU(N)$ Yang-Mills theory and $U(1)$ lattice gauge theory. 
Our most important finding about  $U(1)$ lattice theory is that if one treats  the system along the same line as in the so called non-interacting  instanton gas analysis \cite{Callan:1977gz}, the theory possess the very same ``infrared embarassment'' problem \cite{Coleman:1985rnk, Polyakov:2004vp}.  
However, incorporating the interactions rigorously,  the lattice theory can be dualized to an abelian Higgs model (with a magnetic scalar field) \cite{Peskin:1977kp,Stone:1978mx}. 
In other words, we proved that the so called ``infrared embarassment'' problem, in this set-up, is actually fake, a consequence of ignoring long-range interactions between magnetic current lines. 

Then, we proved that the long-distance field of an instanton can be reproduced in terms of $(N^2-1)$ $U(1)$ self-dual magnetic dipoles, which can be assembled into an $SU(N)$ gauge field.  
These dipoles, by the ADHM construction, are reproduced by $4N-4$ parameters of centered moduli space ${\mathcal M}_{\rm c}$.  
This connects rather mathematical formalism of ADHM  \cite{Dorey:2002ik, Atiyah:1978ri}  to the physical notion of  non-abelian magnetic dipole moment  ${\bm M}_{\mu \nu}  =  M^{a}_{\mu \nu} T^a$. 

Certain aspects of our construction still remain open. 
We did connect bilinears of the ADHM parameters to non-abelian dipole moments, but we did not yet construct a full microscopic description of the instanton in terms of  $w_{u \dot \alpha } \in {\cal M}_{\rm c}$.  
This feature was originally what we were after, and  we are still exploring it.  

In abelian lattice gauge theory, a clear description of the mapping between the original electric description and the dual magnetic description is present.  
It is very little appreciated that this beautiful lattice duality is just the minimal bosonic truncation of the duality in the work of Seiberg and Witten \cite{Seiberg:1994rs},
where the original ${\cal N}=2$  Seiberg-Witten theory at the monopole point in moduli space is mapped to ${\cal N}=2$ SQED where the photon is the dual of the original photon, and the scalar is nothing but the monopole. 
Of course, in  ${\cal N}=2$   theory, the (dual photon + monopole scalar) is completed to their respective  full ${\cal N}=2$ multiplets.  
One of our hopes is that perhaps, in the same spirit as in $U(1)$ lattice gauge theory, we may construct a dual field theory description for the internal structure of instanton. 
Although a non-abelian electric-magnetic duality is a hard problem, we note that,  in the long-range field of instanton,  $[A_{\mu}, A_{\nu}]  \ll \partial_{\mu} A_{\nu} -   \partial_{\nu} A_{\mu}$.   (The former is $\rho^2/|x|^6$ and the latter is  $2/|x|^4$.)
  And what matters in building the matrix  field  $F_{\mu \nu}= F_{\mu \nu}^a T^a$  of an instanton at long distances is that essentially the $F_{\mu \nu}^a$ behave like the strengths of abelian fields.

Clearly, we have high hopes from the formalism we are trying to build up.  Perhaps, a complete and unambiguous  theory of confinement and mass gap on ${\mathbb R^4}$  will ultimately arise. 

\acknowledgments 
M.\"U. thanks Larry Yaffe and Liviu I. Nicolaescu for correspondence on 
counting of the non-backtracking loops on the lattice.  M. \"U. is supported by U.S. Department of Energy, Office  of Science, Office of Nuclear Physics under Award Number DE-FG02-03ER41260.

\appendix

\section{Lattice notation}
Sites of the original lattice are denoted by $x,x',\ldots,$ sites of the dual lattice by $z,z',\ldots .$
The standard orthonormal basis vectors are denoted by $\hat 1, \hat 2,\ldots,\hat d$.
The forward and backward unit difference operators are denoted by $\Delta_\mu$ and $\nabla_\mu$, respectively; i.e., 
\begin{equation}
    \begin{aligned}
    \Delta_\mu f(x) &= f(x+\hat\mu) - f(x) \\
    \nabla_\mu f(x) &= f(x) - f(x-\hat\mu)
    \end{aligned}
\end{equation}
Contracting all indices of a $p$-index antisymmetric tensor field on the original lattice with $p$ indices on the $\epsilon$ tensor produces a $(d-p)$-index antisymmetric tensor field on the dual lattice (the ``Hodge dual'' tensor) according to the rule
\begin{equation}
    \tilde f_{\mu_{p+1}\ldots\mu_d}(\tilde x) = f_{\mu_1\ldots\mu_p}(x) \epsilon_{\mu_1\ldots\mu_p\mu_{p+1}\ldots\mu_d} / p!
    \label{eq:lattice-hodge-dual}
\end{equation}
where $\tilde x = x+\tfrac12 \hat 1 + \ldots \tfrac12 \hat d$, which is a site on the dual lattice.%
\footnote{
One can view a $p$-index antisymmetric tensor field on a lattice as a function on the $p$-cells of the lattice, where $f_{\mu_1\ldots\mu_p}(x)$ is the value on the $p$-cell whose vertices are given by all $x + a_1 \hat \mu_1 + \ldots + a_p\hat\mu_p$ with each $a$ equal to $0$ or $1$. 
The ``Hodge dual'' tensor is then a function on $(d-p)$-cells of the dual lattice, where $\tilde f_{\mu_{p+1}\ldots\mu_d}(\tilde x)$ is the value on the $(d-p)$-cell whose vertices are given by all $\tilde x + b_{p+1} \hat \mu_{p+1}+\ldots+b_d\mu_d$ with each $b$ equal to $0$ or $-1$.
}
We sometimes write the lattice-site argument of a lattice field as a suffix preceding the tensor indices, e.g.,
\begin{equation}
    f_{x,\mu\nu} = f_{\mu\nu}(x)
\end{equation}

\section{The transmutation of instanton operator: $SU(N)$ } 
In this appendix, we generalize our discussion on the matching of the instanton operators   
for the $SU(2)$ gauge theory given in Section \ref{sec:trans} to $SU(N)$. 

\subsection{Instanton operator on $\R^3 \times S^1$ from monopole operators: $SU(N)$ case } 
In $SU(N)$ gauge theory that remains center-symmetric on the small $\R^3 \times S^1$ regime, 
the gauge dynamics abelianize because of the gauge holonomy potential and long distance theory reduces to $U(1)^{N-1}$ gauge group. In this regime, there are $N$ monopole-instanton constituents of the 4d instanton.  The   monopole operators in this abelianized regime are given by 
\begin{align}
{\cal M}_i (x) =    e^{ - \langle \alpha_i , \phi  (x) \rangle +  i  \langle \alpha_i ,\sigma(x)\rangle} \equiv   e^{ - \langle \alpha_i ,z(x)\rangle}, \qquad z \equiv \phi - i \sigma
\label{monop2}
\end{align} 
Here, $\phi$ consists of the Cartan components of the gauge holonomy field, $\sigma$ consists of the dual photon fields, and $\langle , \rangle$ denotes the inner product on $SU(N)$ weight space. 

In statistical field theory representation of Coulomb gas, the charges  $q_i$ with interactions $\frac{q_i q_j}{4 \pi r}$ are represented by the operators $e^{i q_i \sigma}$, so that the Coulomb interaction follows. To generate the absence of  interaction  between mutually BPS pairs  and the 
doubling of the interaction between BPS and anti-BPS pairs, we must invoke both real and imaginary charges to statistical field theory. In fact, the monopole operator  \eqref{monop2} is manifestation of this idea. It can be viewed as a pair of charges, one purely imaginary and the other purely real (the magnetic charge), for the purpose of statistical field theory: 
 \begin{align} 
{ \rm monopoles:} \qquad &
( i \alpha_i, \alpha_i) \in i \Gamma_{\rm r} \oplus  \Gamma_{\rm r}  \cr
{ \rm antimonopoles:} \qquad  & ( i \alpha_i, -\alpha_i) \in i \Gamma_{\rm r} \oplus  \Gamma_{\rm r} 
 \end{align}
In this way,  we have 
 \begin{align} 
&\langle ( i \alpha_i, \alpha_i) ,  ( i \alpha_j, \alpha_j) \rangle = ( i^2+1)   \langle \alpha_i, \alpha_j \rangle =0, \cr
& \langle ( i \alpha_i, \alpha_i) ,  ( i \alpha_j,  -\alpha_j)\rangle = ( i^2-1)  \langle\alpha_i,  \alpha_j \rangle = -2  \langle\alpha_i, \alpha_j\rangle
\label{int-mon}
 \end{align}
 i.e., mutually BPS pairs do not interact, while BPS and anti-BPS pairs interact via long-range field.\footnote{The charge multiplying the holonomy field is not an electric charge. 
 (At best, it can be characterized as an imaginary electric charge.)  In particular, the magnitude of these two charges are equal to each other, unlike dyons, for which  $Q_e  \propto   e  $ and $Q_m  \propto   1/ e  $ are reciprocal and  satisfy Dirac quantization. Hence, this object is not a dyon.} 
In supersymmetric theories, e.g. ${\cal N}=1$ SYM,  monopoles remain BPS to 
 all orders in perturbation theory. In center-symmetric non-susy theories,  a perturbative mass for the $\phi$ field is generated.   


Let us consider an instanton of fixed size. We assume the size is much larger than compactification scale, hence we know that the instantons fractionates. 
$SU(N)$ instanton as a composite of $N$ monopoles can be written as: 
\begin{align}
{\cal I}  = \prod_{i=1}^{N}  {\cal M}_i = \prod_{i=1}^{N}   e^{-  \langle\alpha_i,z(x_i)\rangle}
\label{montoI}
\end{align}  
Denote the center position of the collection of $N$ monopoles by $x_{\rm c}$, and relative coordinates $\xi_i$, hence, 
\begin{align}
x_{\rm c}= \frac{1}{N} \sum_{i=1}^{N} x_i,  \qquad \xi_i = x_i -x_{ \rm c}
\end{align}  
Note that $x_i, i=1,\ldots  N$ are not only the position of 3d monopole constituents, they are also moduli parameters of instanton. Each monopoles also has a U(1) angular moduli, building $3N + N=4N$ moduli of the instanton.  \eqref{monop2}, \eqref{montoI} and \eqref{int-mon} make manifest the fact that the $x_i \in \R^3 $ are moduli of the instanton.

Since instanton is of fixed size,  we can expand the the  field $z(x_i)$ around the center position $x_{ \rm c}$ as   $z( x_{\rm c} +\xi_i) = z( x_{ \rm c}) +  (\xi_i. \nabla) z (x_{\rm c}) + \cdots $. 
Note that \eqref{montoI} is magnetically neutral because  $\sum_{i=1}^{N} \alpha_i  =0$, hence, the instanton does not carry a net magnetic charge. Thus, we must view it some sort of a dipole.  Below, we describe the sense in which instanton is a dipole on small $\R^3 \times S^1$.  Expanding the instanton operator around the $x_{ \rm c}$, we find: 
\begin{align}
{\cal I} 
&= \underbrace{ \prod_{i=1}^{N}   e^{-  \langle \alpha_i ,z(x_{\rm c} )\rangle}   }_{=1}
 \prod_{i=1}^{N} e^{-  \langle \alpha_i ,  (\xi_i. \nabla)z(x_{\rm c} )\rangle} 
 = \prod_{i=1}^{N} e^{-  (\xi_i. \nabla)(z_i - z_{i+1})(x_{\rm c} )} =  \prod_{i=1}^{N} e^{-   (\xi_i -\xi_{i-1})   \nabla  z_i (x_{\rm c} )} \cr
& = \prod_{i=1}^{N} e^{-   \vec d_i \cdot \vec \nabla  z_i (x_{\rm c} )} = \prod_{i=1}^{N} e^{-   \vec d_i   \vec \nabla  \phi_i (x_{\rm c} ) + i  \vec d_i   \vec \nabla   \sigma_i (x_{\rm c} ) }
\label{Ins-mon}
\end{align}  
It is as if we have $N$ real and  $N$ imaginary dipoles, associated with each component,
and  satisfying  a constraint 
\begin{align}
    ( i \vec d_i, \vec d_i) \in (i \R^3, \R^3) , \;\;  i=1, \ldots, N,  \qquad \sum_{i=1}^{N}  \vec d_i = 0
    \label{dipole3rep}
\end{align}
 We find this representation useful, because it makes the 
 vanishing of the instanton-instanton interaction and the dipole-dipole nature of the instanton-antiinstanton interactions (now, this is 3d dipoles because of the dimensional reduction) manifest. 
 
Moreover, \eqref{dipole3rep} tells us that we can repackage the $3N$ position type moduli of 
$N$ monopoles as $3N-3$ dipoles (because of the one constraint), plus $x_{\rm c}$ center-position moduli. This representation will match directly to the dimensional reduction of the instanton operator from 4d, which we explore below.

\subsection{From instanton operator on $\R^4$ to the one in $\R^3 \times S^1$: $SU(N)$ case} 
We expressed the instanton operator in 4d  as:
\begin{align}
 I &= {\rm exp}\left\{-  {\rm tr} (w \bar \sigma_{mn} \bar w F_{mn}) \right\} \cr
 & =  {\rm exp}\left\{-  {\rm tr} (w  \tau^a  \bar w  (E^a + B^a) ) \right\}, \qquad a= 1,2,3
\end{align} 
Here, $w_{u, \dot \alpha}, u=1, \ldots, N,  \dot \alpha=1,2 $  are parameters entering ADHM construction.  Let us denote it in terms of two $N$ component complex vectors:
\begin{align}
 w =  \begin{bmatrix}
         \a_{1} &  \b_{1}   \\
          \a_{2} &  \b_{2}   \\
    \vdots & \vdots  \\
     \a_{N} &  \b_{N}   \\
    \end{bmatrix}  \qquad   \bar w =  \begin{bmatrix}
         \bar \a_{1} &  \bar \a_{2}  & \hdots &   \bar \a_{N}    \\
          \bar \b_{1} &  \bar \b_{2}  & \hdots &   \bar \b_{N} 
    \end{bmatrix}  
 \end{align}
which  satisfy the following  constraints:
\begin{align}
  &   \wbar_{u}^{\alphadot} w_{u\betadot} = 
  \begin{pmatrix}
         |\a|^2 &  \a^{\dagger} \b  \\
      \b^{\dagger} \a  & |\b|^2
    \end{pmatrix}=  
    \begin{pmatrix}
         \rho^2 &  0  \\
        0 & \rho^2
    \end{pmatrix} 
  \label{constraints}
\end{align} 
Imposing the  ADHM constraints, and moding with the 
$U(1)$ redundancy  (which is a $U(k)$ redundancy in $k$ instanton construction \cite{Dorey:2002ik},   they give a  $4N-4$ dimensional parametrization of the centered moduli space ${\cal M}_{\rm c}$.   

Given this, we can express the self-dual non-abelian magnetic dipole moments  of the instanton  $w \bar\sigma_{\nu\alpha} w^\dag=  \sum_{i=1}^3 w \tau^i w^\dag \;  \bar \eta^i_{\mu \nu}$  in terms of ADHM data as:
\begin{align}
     w \tau_{1} \wbar = &  \a \otimes  \b^{\dagger}  + \b \otimes  \a^{\dagger}  \cr 
     w \tau_{2} \wbar =&  -i (\a \otimes  \b^{\dagger}  - \b \otimes  \a^{\dagger}  )  \cr 
       w \tau_{3} \wbar  =&   \a \otimes  \a^{\dagger}  -  \b \otimes  \b^{\dagger} 
          \label{moments2}
\end{align}
Note that the ADHM constraints is equivalent to the tracelessness conditions of these matrices.

When the dynamics abelianize on $\R^3 \times S^1$, only $U(1)^{N-1}$ subgroup of $SU(N)$ remains gapless.    As a result,  at long distances, only Cartan components of the gauge field survive. Thus, in this regime, the instanton operator becomes  (using abelianization and the 3d abelian duality for the magnetic field, as in our $SU(2)$ discussion)  becomes 
\begin{align}
 I& =  {\rm exp}\left\{-   ((w  \vec \tau  \bar w)_{uu}  (   \nabla \phi   -    i \nabla \sigma )_{u}) \right\}, \qquad   \; u=1, \ldots N
\end{align} 
We clearly see that this is identical  with the instanton operator obtained from the monopole-instanton description \eqref{Ins-mon}. Hence, we identify the 3d magnetic dipole moments in terms of 4d ADHM moduli parameters as:
\begin{align}
\vec d_u  = \vec  \xi_u -   \vec \xi_{u-1}  = L^{-1} (w \vec \tau \wbar)_{uu}  =  L^{-1}(  2 {\rm Re } ( \a_{u} \bar \b_{u}), 2 {\rm Im } ( \a_{u} \bar \b_{u}), 
 |\a_{u}|^2 -  | \b_{u}|^2 )
     \label{moments}
\end{align}
The modulus of  the dipole moments, for each $u$, is given by: 
\begin{align}
 |\vec d_u| =  
   L^{-1} \rho_u^2 = L^{-1}  ( |\a_{u}|^2 +   |\b_{u}|^2), \qquad 
     \label{dipolelengths}
\end{align}

The sum of the modulus of 3d dipole moments is noting but the 
 size modulus  of the 4d instanton:
\begin{align}
&L \sum_{u=1}^{N}  |\vec d_u| =\sum_{u=1}^{N}    |\a_{u}|^2 +   |\b_{u}|^2 =  \a^{\dagger}\a +    \b^{\dagger}\b= 2 \rho^2,    \cr
&  2 \rho^2 =  L \sum_{u=1}^{N}   | \vec \xi_u -   \vec \xi_{u-1}|,  \qquad   \xi_{0} \equiv  \xi_{N}
     \label{dipolelengths}
\end{align}
We also observe that there is a generalization of the Hopf map in the $SU(2)$ story to the present case. The $N-1$ of the  angular moduli of the $N$-monopoles are realized in terms of fibres of the Hopf maps,  $S^3_u \rightarrow S^2_u $. 

To wrap up, we see that the dimensional reduction of the 4d instanton operator we constructed in the present work matches precisely with the instanton operator obtained by taking the product of monopole-instanton operators on $\R^3 \times S^1$. This provides a physical way to interpret the bilinears appearing in the ADHM construction in terms of simple magnetic dipole moments of the monopole-instantons.

\bibliographystyle{utphys}
\bibliography{./QFT-Mithat.bib,refs}
  
\end{document}